\documentclass[namedreferences]{SolarPhysics}
\usepackage[optionalrh]{spr-sola-addons} 
\usepackage{graphicx} 
\usepackage{color} 
\usepackage{url} 

\newcommand{\beq}{\begin{equation}}
\newcommand{\eeq}{\end{equation}}
\newcommand{\beqa}{\begin{eqnarray}}
\newcommand{\eeqa}{\end{eqnarray}}

\begin{document}


\begin{opening}

\title{Numerical simulations of Magnetoacoustic-Gravity Waves in the Solar Atmosphere}

\author
{
K.~   \surname{Murawski}$^{1}$\sep
A.~K.~ \surname{Srivastava}$^{2}$\sep
J.~A.~\surname{McLaughlin}$^{3}$\sep
R.~   \surname{Oliver}$^{4}$\sep
}
    
\runningauthor{Murawski et al.}
\runningtitle{Magnetoacoustic-Gravity Waves in the Solar Atmosphere}
\institute
{
$^{1}$Group of Astrophysics,
             UMCS, ul. Radziszewskiego 10, 20-031 Lublin, Poland. email: \url{kmur@kft.umcs.lublin.pl}\\
$^{2}$ Aryabhatta Research Institute of Observational Sciences (ARIES), Nainital, India.
                     email: \url{aks@aries.res.in}\\
$^{3}$School of Computing, Engineering and Information Sciences, 
Northumbria University, Newcastle Upon Tyne, NE1 8ST, UK.
                     email: \url{james.a.mclaughlin@northumbria.ac.uk}\\
$^{4}$Departament de F\'isica, Universitat de les Illes Balears, 07122, Palma de Mallorca, Spain.
                     email: \url{ramon.oliver@ uib.es}
}

\begin{abstract}
{
We investigate  the excitation of magnetoacoustic-gravity waves generated from localized pulses in the gas pressure as well as in vertical component of velocity. 
These pulses are initially launched at the top of the solar photosphere that is permeated by a weak magnetic field. 
We investigate three different configurations of the background magnetic field lines: horizontal, vertical and oblique to the gravitational force. 
We numerically model magnetoacoustic-gravity waves by implementing a realistic (VAL-C) model of solar temperature. We solve two-dimensional ideal magnetohydrodynamic equations numerically with the use of {\bf{the}} FLASH code to simulate the dynamics of {\bf{the}} lower solar atmosphere. The initial pulses result in shocks at higher altitudes. Our numerical simulations reveal that a small-amplitude initial pulse can produce magnetoacoustic-gravity waves, which are later reflected from the transition region due to the large temperature gradient. 
The atmospheric cavities in the lower solar atmosphere are found to be the ideal places that may act as a resonator for various oscillations, including their trapping and leakage into the higher atmosphere. Our numerical simulations successfully model the excitation of such wave modes, 
their reflection and trapping, as well as the associated plasma dynamics.
} 
\end{abstract}
\keywords{Waves, Magnetic fields, Corona, Granulation}
\end{opening}





%
%
\section{Introduction}
The complicated magnetic field configuration of the Sun plays a key role in various types of
dynamical plasma processes in its atmosphere, including all the significant plasma dynamics of
the lower solar atmosphere. 
The resulting  magnetic structures channel energy from the photosphere into the upper atmosphere, in form of magnetohydrodynamic (MHD) waves, 
and such waves  experience mode conversion, resonances, trapping and reflection which results in the complicated dynamical processes in the lower solar atmosphere, 
the details of which depend on the plasma properties as well as strength of the magnetic field. 
The complex magnetic field and plasma structuring in the lower solar atmosphere support the excitation of various kinds of MHD waves, and their propagation, 
reflection and trapping has been studied extensively both on theoretical and observational grounds 
(e.g. Murawski et al., 2011; Srivastava \& Dwivedi, 2010;
Srivastava, 2010; Fedun et al., 2009; Srivastava et al., 2008; Hasan et al., 2005;
McAteer et al., 2003; Gruszecki et al., 2011, and references therein).
The evolving magnetic fields of the lower solar atmosphere also lead to transient processes 
across a wide range of spatial-temporal scales in form of the eruption and associated phenomena. 
For example, various types of solar jets are formed at short spatial-temporal scales which play a  significant role  
in mass and energy transport and also  couple the various layers of the solar atmosphere 
(e.g. Shibata et al., 2007; Katsukawa et al., 2007; Srivastava \& Murawski, 2011, and references
therein).
In addition, the magnetic activity and injections of helicity into the lower solar atmosphere result in large-scale eruptive phenomena, 
including solar flares and coronal mass ejections (CMEs) in the  outer part of the magnetized solar atmosphere 
(e.g. Srivastava et al., 2010; Shibata \& Magara, 2011; Zhang et al., 2012, and references therein).
Therefore, the coupling of the complex magnetic field in various layers of the Sun due to waves and transients is one of the most significant areas of contemporary solar research.

In the quiet-Sun magnetic networks, cavities are important locations where the magnetic fields are sufficiently inclined due to their well evolved horizontal components. 
They are formed over the granular cells in form of the field-free regions due to the transport of plasma at their boundaries and overlaid by bipolar magnetic canopies. 
The vertical magnetic fields, however, reside mostly in the core of such magnetic networks (Schrijver \& Title, 2003; Centeno et al.,
2007). 
Such magnetic cavity-canopy systems are thought to be ideal resonators of the various MHD waves that can be trapped in the cavity, 
as well can also leak in form of the magnetoacoustic-gravity waves upward through the core of such magnetic networks. 
It is thought that the field-free cavity regions underlying the bipolar canopy can trap the high-frequency acoustic oscillations, 
and the low-frequency components may leak into the higher atmosphere in form of magnetoacoustic-gravity waves 
(Kuridze et al., 2008; Srivastava et al., 2008; Srivastava, 2010). 
Therefore, such magnetic structures in the lower solar atmosphere are the regions 
that may play an important role in the wave filtering 
(e.g. McIntosh \& Judge, 2001; Krijger et al., 2001; McAteer et al., 2002; Vecchio
et al., 2007, 2009; Srivastava, 2010, and references therein).

Alongside recent high-resolution observations of MHD waves in the lower solar atmosphere, 
extensive efforts have been {\bf{made}} in the area of analytical and numerical modeling of such waves: 
Fedun et al. (2009)
have {\bf{investigated}} the 3D numerical modeling of the coupled slow and fast magnetoacoustic wave propagation in the lower solar atmosphere; 
and recently 
Fedun et al. (2011a) have {\bf{reported the first}} numerical results of the frequency filtering of torsional Alfv\'en waves in the chromosphere. 
Apart from general numerical modeling of the waves in the lower solar atmosphere, models have also {\bf{investigated}} the acoustic wave spectrum in the localized magnetic structures of the lower solar atmosphere, e.g., magnetic cavity-canopy system (e.g. Kuridze et al., 2008; Srivastava et al.,
2008; Kuridze et al., 2009, and references cited therein).
  
It is also noteworthy that Bogdan et al. (2003), Fedun et al. (2009) and Fedun
et al. (2011a)
discussed in detail the excitation, propagation and conversion of magnetoacoustic waves in a realistic 3D MHD simulation. 
However, in these references, the waves driven by a periodic driver were discussed, whereas we numerically simulate the excitation 
of magnetoacoustic-gra\-vi\-ty waves 
generated by pulses in the gas pressure and vertical component of velocity, mimicking an isolated solar granule. 
We aim to investigate and understand this simpler (albeit complex enough) system 
before tackling the more realistic, multiple granule system. Our philosophy is to build up our models incrementally, 
with a clear focus on the underlying physical processes at each step. 

In this paper, we  investigate  the excitation of magnetoacoustic-gravity waves generated from localized pulses in the gas pressure as well as 
in vertical component of velocity, modelling the effect of an isolated solar granule. 
These pulses are initially launched at the top of the solar photosphere that is permeated by a weak magnetic field. 
We investigate three different configurations of the background magnetic field lines: 
vertical, 
horizontal, 
and oblique 
to the gravitational force. We aim to show that small amplitude perturbations that are associated with such a granule are able to trigger large amplitude, 
complicated oscillations in the solar corona, which exhibit periodicities within the detected range of $3-5$ min. 

The structure of the paper is as follows: In 
Sect.~\ref{SECT:NUM_MODEL} we describe the numerical model. 
We report the numerical results in Sect.~\ref{sect:results} and present the discussion and conclusions in {\bf{Sect.~\ref{SECT:DISS} }}.
%

%
%
\section{Numerical model}\label{SECT:NUM_MODEL}
We consider a gravitationally-stratified solar atmosphere that is described by the ideal two-dimensional (2D) MHD equations:
\beqa
\label{eq:MHD_rho}
{{\partial \varrho}\over {\partial t}}+\nabla \cdot (\varrho{\bf V})&=&0\, ,
\\
\label{eq:MHD_V}
\varrho{{\partial {\bf V}}\over {\partial t}}+ \varrho\left ({\bf V}\cdot \nabla\right ){\bf V} &=& -\nabla p+ \frac{1}{\mu}(\nabla\times{\bf B})\times{\bf B} +{\varrho{\bf {g}}\,} ,
\\
\label{eq:MHD_p}
{\partial p\over \partial t} + \nabla\cdot (p{\bf V}) &=& (1-\gamma)p \nabla \cdot {\bf V}\, ,
\\
\label{eq:MHD_B}
{{\partial {\bf B}}\over {\partial t}}&=& \nabla \times ({\bf V}\times{\bf B})\, , 
\hspace{3mm}
\nabla\cdot{\bf B} = 0\, .
\eeqa
Here ${\varrho}$ is mass density, ${\bf V}$ is the flow velocity, ${\bf B}$ is the magnetic field, 
$p = {k_{\rm B}} \varrho T / m$ is the gas pressure, $T$ is the temperature, $\gamma=5/3$ is the adiabatic index, 
${\bf g}=(0,-g,0)$ is the gravitational acceleration, {\bf{where}} $g=274$ m s$^{-2}$, $m$ is the mean particle mass 
and $k_{\rm B}$ is Boltzmann's constant. 
{\bf { Throughout this paper, we use the Cartesian coordinate system with the vertical axis denoted by $y$ and 
the horizontal axis $x$. 
Henceforth, we assume that the medium is invariant along the $z$-direction with $\partial/\partial z = 0$ and 
set the $z$-components of plasma velocity and magnetic field equal to zero, i.e. $V_{\rm z}=0$ and $B_{\rm z}=0$. 
The latter assumption removes Alfv\'en waves from the system but still allows  magnetoacoustic-gravity waves to propagate freely. 
}}
%
%
%
\subsection {Equilibrium configuration}
\begin{figure}[!h]
\begin{center}
\includegraphics[scale=0.34, angle=0]{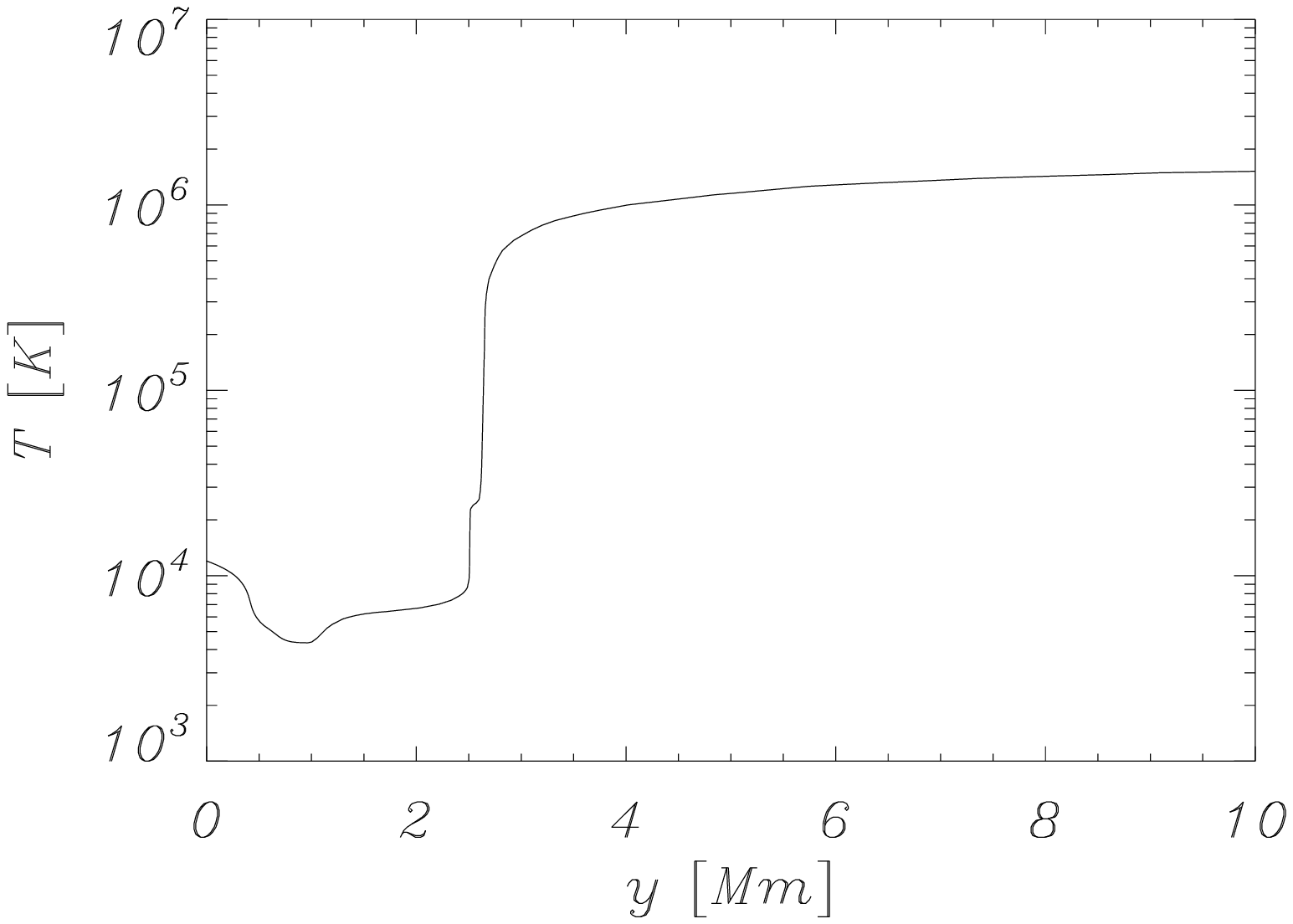}
\includegraphics[scale=0.34, angle=0]{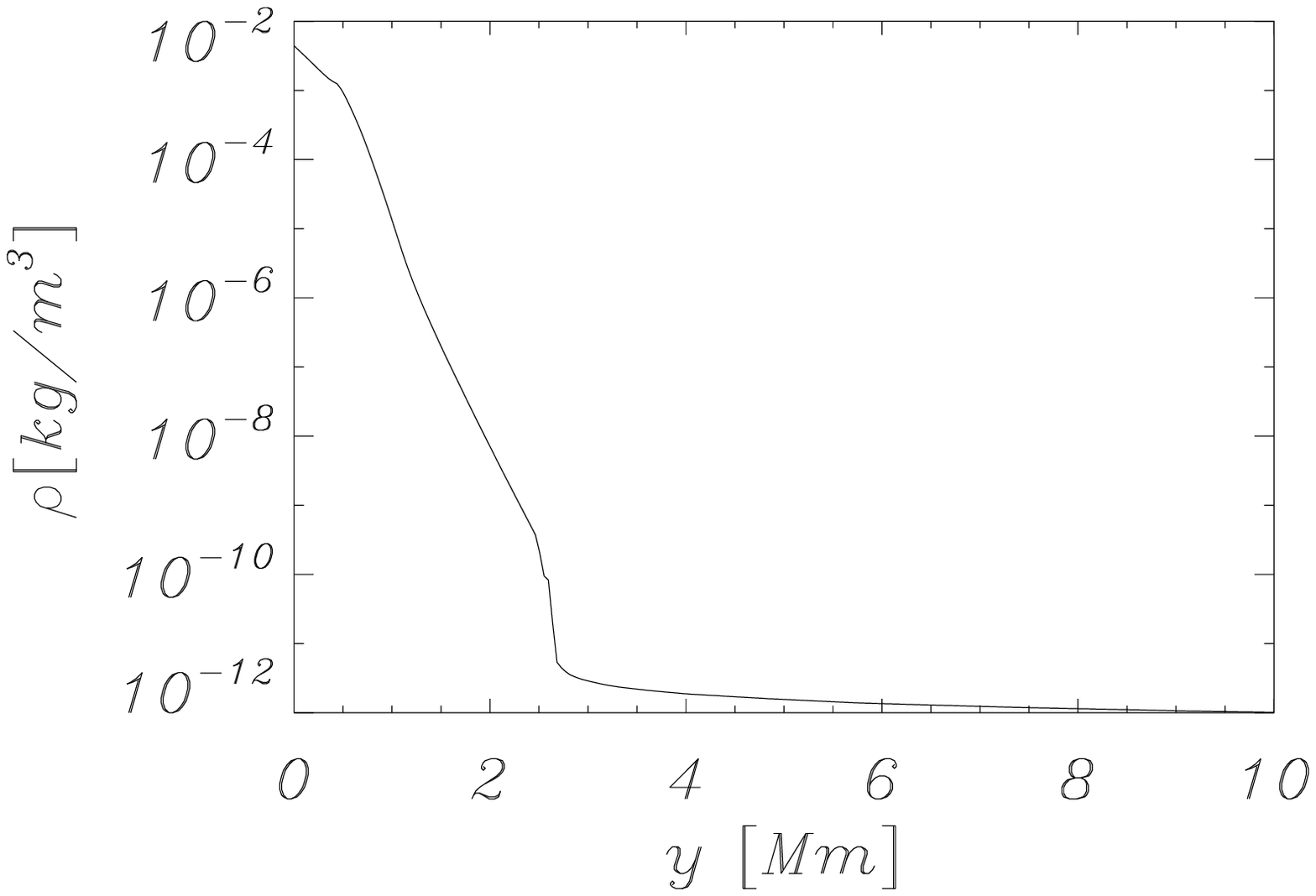}\\
\includegraphics[scale=0.35, angle=0]{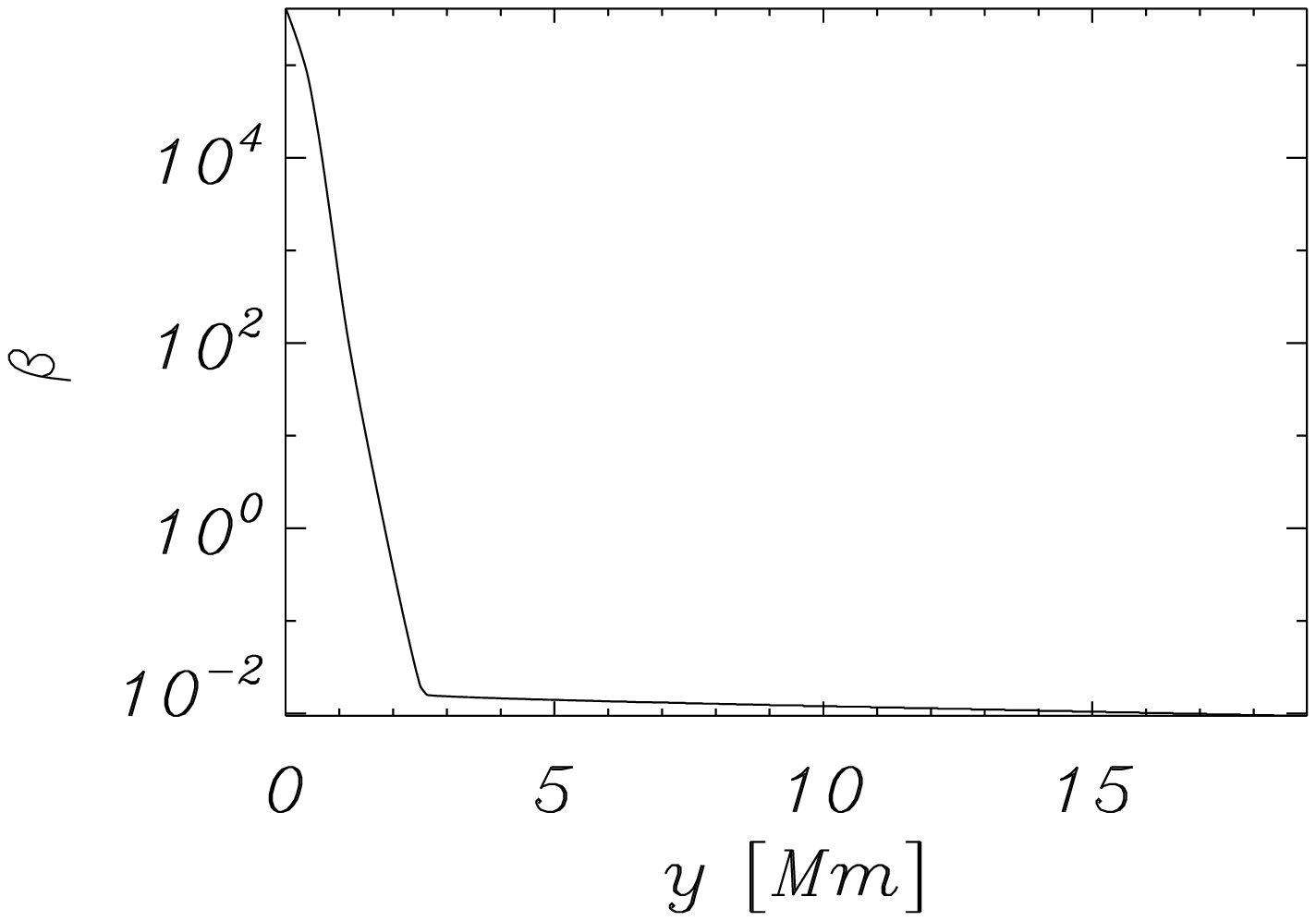}
\caption{\small
Equilibrium profile of solar temperature (top-left), mass density (top-right),  
and the plasma $\beta$ (bottom). 
}
\label{fig:initial_profile}
\end{center}
\end{figure}

We assume that at its equilibrium the solar atmosphere is static (${\bf V}_{\rm e}={\bf 0}$) and threaded by a straight magnetic field, 
\beq\label{eq:bb}
{\bf B} = B_{\rm 0} {\bf\hat s}\, ,
\eeq
where ${\bf\hat s}$ is a unit vector that is either vertical, horizontal or oblique. 
We choose $B_{\rm 0}$ 
by requiring that at the reference point $(0,10)$ Mm, the Alfv\'en speed
\beq\label{eq:c_A}
c_{\rm A}(y) = \frac{B_{\rm 0}}{\sqrt{\mu \varrho_{\rm e}(y)}}
\eeq
and sound speed
\beq\label{eq:c_s}
c_{\rm s}(y) = \sqrt\frac{\gamma p_{\rm e}(y)}{\varrho_{\rm e}(y)}
\eeq
satisfy the following constraint: $c_{\rm A}(y=10\, {\rm Mm}) = 10\, c_{\rm s}(y=10\, {\rm Mm})$. 
This constraint reproduces typical conditions in the solar corona where, typically, $c_{\rm s}=0.1$ Mm s$^{-1}$ and $c_{\rm A}=1$ Mm s$^{-1}$. 
As a result, the solar corona is magnetically dominated with $B_{\rm 0}\simeq 14.5 \times 10^{-4}$ T. The plasma $\beta$, 
\beq
\beta(y) = \frac{\gamma}{2} \frac{c_{\rm s}^2(y)}{c_{\rm A}^2(y)}\, , 
\eeq
in the solar corona attains a value of $\beta(y=10\,{\rm Mm}) = 0.012$. It grows slowly with depth within the chromosphere and below $y=2$ Mm reaches abruptly a value of 
$\beta(y=0\,{\rm Mm}) \simeq 4 \times 10^{5}$ at the bottom of the photosphere (Fig.~\ref{fig:initial_profile}, bottom). This large value of $\beta$ evidences that in these low regions of the solar atmosphere 
the effect of magnetic field is negligibly small. 

The straight magnetic field of Eq.~(\ref{eq:bb}) is a simplified model of curved magnetic field lines which are located in low regions of the solar atmosphere. 
However, as the plasma $\beta$ grows fast with depth below $y=2.5$ Mm reaching large values there (see Fig.~\ref{fig:initial_profile}, bottom) 
we actually model the very weakly-magnetized low layers of the atmosphere, 
which do not correspond to flux-tubes and sunspots. 
Therefore, in this first approximation, the curved magnetic field {\bf{lines}} can be replaced by a straight magnetic field, and our models are justified. 

In Eqs.~(\ref{eq:c_A}) and (\ref{eq:c_s}), $\varrho_{\rm e}(y)$ and $p_{\rm e}(y)$ 
denote equilibrium mass density and gas pressure, respectively. 
%
%
They are specified by the hydrostatic constraint which in the context of Eq.~(\ref{eq:bb}) states that the pressure gradient is balanced by the gravity force, 
\begin{equation}
\label{eq:p}
-\nabla p_{\rm e} + \varrho_{\rm e} {\bf g} = {\bf{0}}\, .
\end{equation}
%
%
%
With the ideal gas law and the $y$-component of Eq.~(\ref{eq:p}), we arrive at 
\beqa
\label{eq:pres}
p_{\rm e}(y)=p_{\rm 0}~{\rm exp}\left[ -\int_{y_{\rm r}}^{y}\frac{dy^{'}}{\Lambda (y^{'})} \right]\, ,\hspace{3mm}
\label{eq:eq_rho}
\varrho_{\rm e} (y)=\frac{p_{\rm e}(y)}{g \Lambda(y)}\, ,
\eeqa
where
\begin{equation}
\Lambda(y) = \frac{k_{\rm B} T_{\rm e}(y)} {mg}
\end{equation}
is the pressure scale-height, and $p_{\rm 0}$ denotes the gas pressure at the reference level that we choose in the solar corona at $y_{\rm r}=10$ Mm.

We adopt an equilibrium temperature profile $T_{\rm e}(z)$ for the solar atmosphere that is close to the VAL-C atmospheric model 
of Vernazza et al. (1981)
: see Fig.~\ref{fig:initial_profile}, left-top panel. 
Note that $T_{\rm e}$ attains a value of about $5700$ K at the top of the photosphere which corresponds to $y=0.5$ Mm. At higher altitudes $T_{\rm e}(y)$ falls off until 
it reaches its minimum of $4350$ K at the altitude $y\simeq 0.95$ Mm. Higher up, $T_{\rm e}(y)$ grows gradually with height up to the transition region which is located at $y\simeq 2.7$ Mm. 
Here $T_{\rm e}(y)$ experiences a sudden growth up to the coronal value of $1.5$ MK at $y=10$ Mm. 
Then with Eq.~(\ref{eq:pres}) we obtain the corresponding gas pressure and mass density profiles. 
Both $p_{\rm e}(y)$ (not shown) and $\varrho_{\rm e}(y)$ (Fig.~\ref{fig:initial_profile}, right-top panel) experience a sudden fall-off from photosphere to coronal values at the transition region. 
%
%
%
\subsection{Initial Conditions}
{\bf{ 
At $t=0$ s, we initially perturb the equilibrium impulsively using localized Gaussian pulses simultaneously in the gas pressure and vertical component of velocity, viz.,}
}
\beqa\label{eq:perturb}
p(x,y,t=0) = p_{\rm e}(y) + 
A_{\rm p} f(x,y)\, ,\hspace{4mm} V_{\rm y}(x,y,t=0) = A_{\rm v} f(x,y)\, , \\
f(x,y) = \exp\left[ -\frac{(x-x_{\rm 0})^2}{{w_{\rm x}}^2}-\frac{(y-y_{\rm 0})^2}{{w_{\rm y}}^2}\right]\, .
\eeqa
{\bf 
Here $A_{\rm p}$ and $A_{\rm v}$ are the amplitudes of the perturbations, $(x_{\rm 0},y_{\rm 0})$ is their initial position
and $(w_{\rm x}, w_{\rm y})$ denotes their widths along the $x$- and $y$-directions. 
}
We set and hold fixed {\bf{$x_{\rm 0}=0$}} Mm, $y_{\rm 0}=0.5$ Mm, $w_{\rm x}=0.5$ Mm, $w_{\rm y}=0.5$ Mm, 
$A_{\rm p} = 0.02\,  p_{\rm e}(y_{\rm 0})$ and $A_{\rm v}=0.2\times 10^{-3}$ Mm s$^{-1}$. 
{\bf 
These magnitudes of $A_{\rm v}$ and $A_{\rm p}$ represent the recently detected flow and temperature in solar granulation (Baran, 2011).
}
We separately consider three {\bf orientations for our unidirectional} equilibrium magnetic field lines: (a) horizontal, (b) vertical, (c) oblique to the gravitational force. 
\section{Numerical results}\label{sect:results}
\begin{figure}
\begin{center}
\includegraphics[scale=0.5, angle=0]{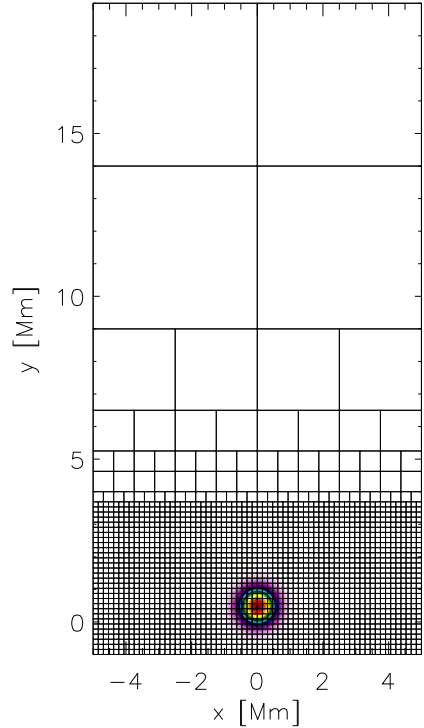}
\caption{\small 
Numerical blocks with their boundaries (solid lines) and the pulse in velocity of Eq.~(\ref{eq:perturb}) (color maps) 
at $t=0$ s for 
{\bf 
the cases of the vertical and horizontal 
equilibrium magnetic fields. 
}
Maximum velocity (red color) corresponds to $0.2\times 10^{-3}$ Mm s$^{-1}$.
}
\label{fig:blocks}
\end{center}
\end{figure}
Equations (\ref{eq:MHD_rho})--(\ref{eq:MHD_B}) are solved numerically using the FLASH code (Lee \& Deane,
2009).
This code implements a second-order unsplit Godunov solver with various slope limiters and Riemann solvers, 
as well as adaptive mesh refinement (AMR). 
{\bf 
The main advantage of using AMR technique is to refine a numerical grid at steep spatial profiles while 
keeping a grid coarse at the places where fine spatial resolution is not essential. Such AMR technique usually introduces 
interpolation errors at different-size numerical cells. These errors can result in vertical flow that, albeit initially small, 
can grow with height to an unacceptable magnitude. A remedy of this inherent phenomenon is to refine the whole region below the transition region, 
which was adopted in our simulations. 
We used the Roe solver and minmod flux limiter, and 
set the simulation box for the horizontal and vertical equilibrium magnetic field cases in the $x$-direction as $-5\le x \le 5$ Mm.
In order to trace plasma structures which extend more horizontally in the case of the oblique magnetic field we use $-2\le x \le 8$ Mm in this case.  
Along the $y$-direction the numerical box was set as $-1 \le y \le 19$ Mm in all cases. 
}
At all boundaries, we fix all plasma quantities to their equilibrium values, which lead only to small numerical reflections of incident wave signals. 
{\bf 
Additionally, we increased the physical domain size and damp incident waves by adopting a coarse numerical grid at the top boundary 
to avoid influence of reflections during the simulation time range. }
As a result, these reflections do not exert any noticeably effect on the dynamics of the system.
In all our studies, we use an AMR grid with a minimum / maximum level of refinement set to $3$ / $7$. 
The initial system of blocks is displayed in Fig.~\ref{fig:blocks}. 
For the vertical and horizontal magnetic fields 
initially, at $t=0$ s the whole simulation region is covered 
by $2666$ blocks, with a similar number of blocks for the oblique magnetic field. 
As every block consists of $8\times 8$ numerical cells, therefore, this number of blocks corresponds to $170624$ numerical cells. 
This results in the finest (poorest) resolution of $\Delta x = \Delta y \simeq 19.5\times 10^{-3}$ Mm ($\Delta x = \Delta y \simeq 625\times 10^{-3}$ Mm) 
in the region below $y=3.7$ Mm (above $y=9$ Mm) 
at $t=0$ s (see Fig.~\ref{fig:blocks}). 
The refinement strategy is based on controlling numerical errors in mass density, which results in excellent resolution of 
steep spatial profiles and greatly reduces numerical diffusion at these locations. 
{\bf The duration of 
a typical numerical run was 
$1500$ s. 
}
Although the numerical simulations have been carried out for $-5$ Mm $\leq x\leq 5$ Mm, $-1$ Mm $\leq y\leq 19$ Mm, 
only the results in the regions of interest are displayed in this work.
%
%
%
\subsection{Vertical equilibrium magnetic field:  ${\bf\hat s}={\bf\hat y}$}\label{sub:vertB}
\begin{figure}
\includegraphics[scale=0.31, angle=0, trim=0cm 0cm 0cm 0cm, clip=true]{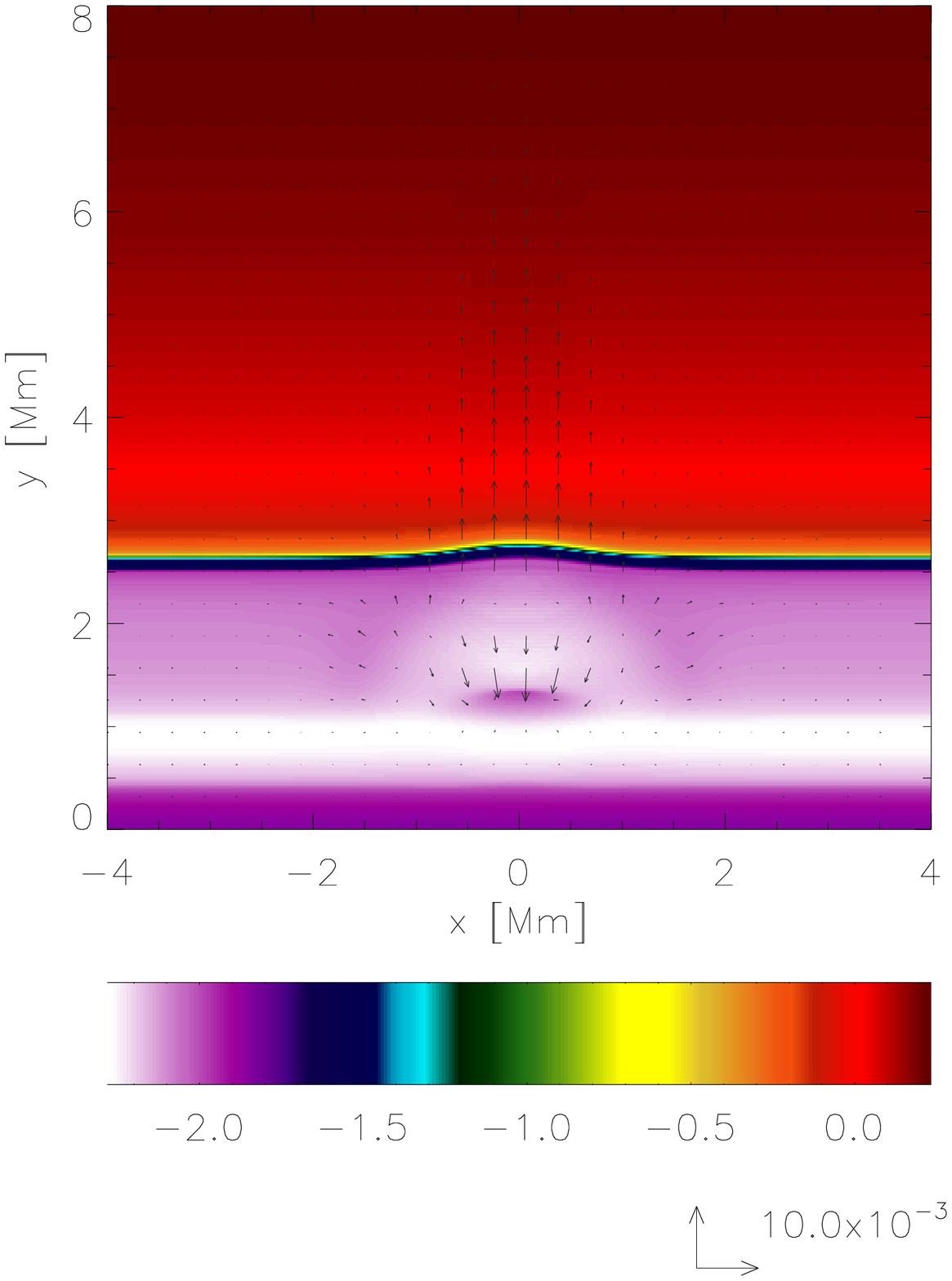}
\includegraphics[scale=0.31, angle=0, trim=0cm 0cm 0cm 0cm, clip=true]{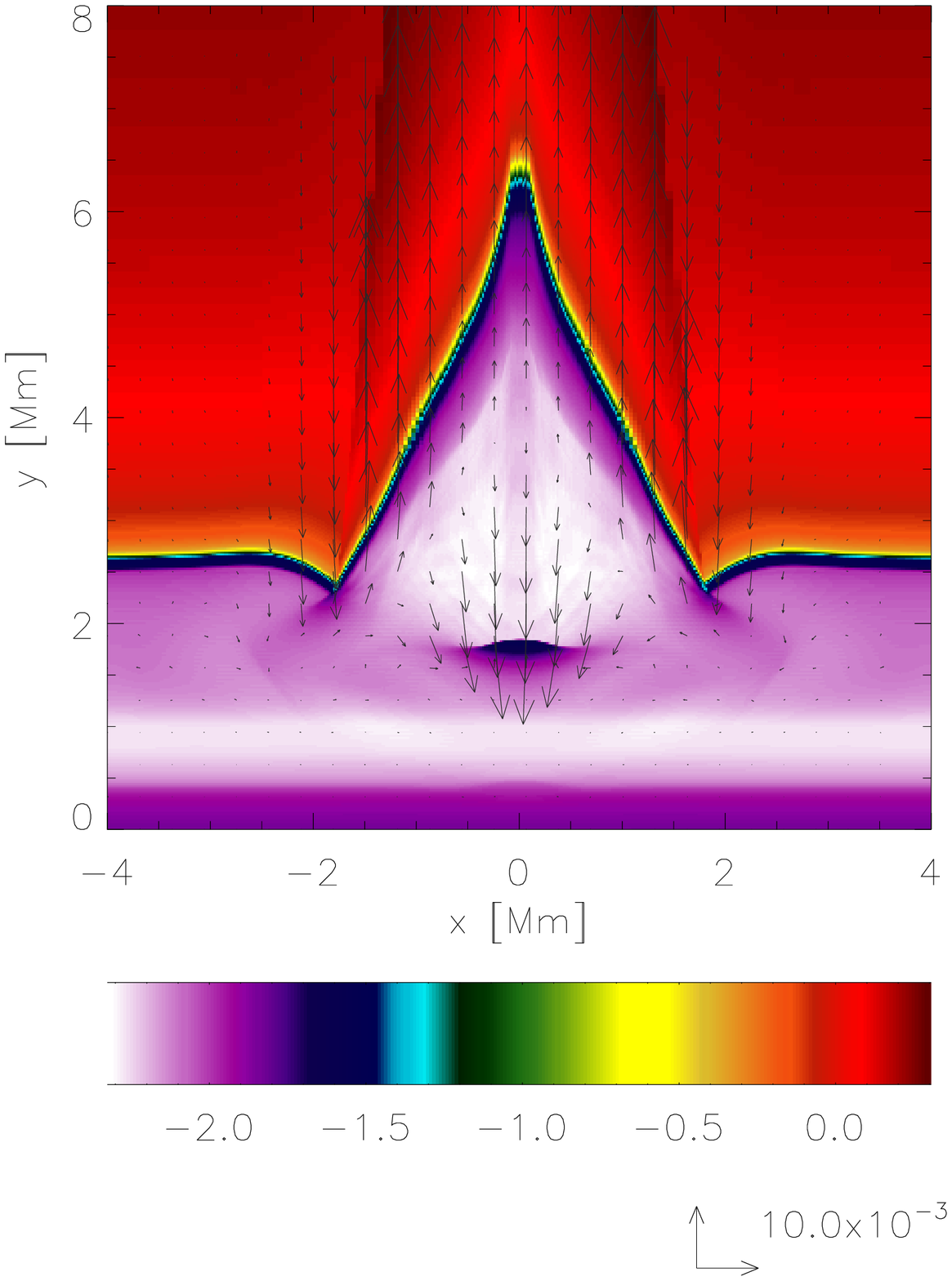}\\
\includegraphics[scale=0.31, angle=0, trim=0cm 0cm 0cm 0cm, clip=true]{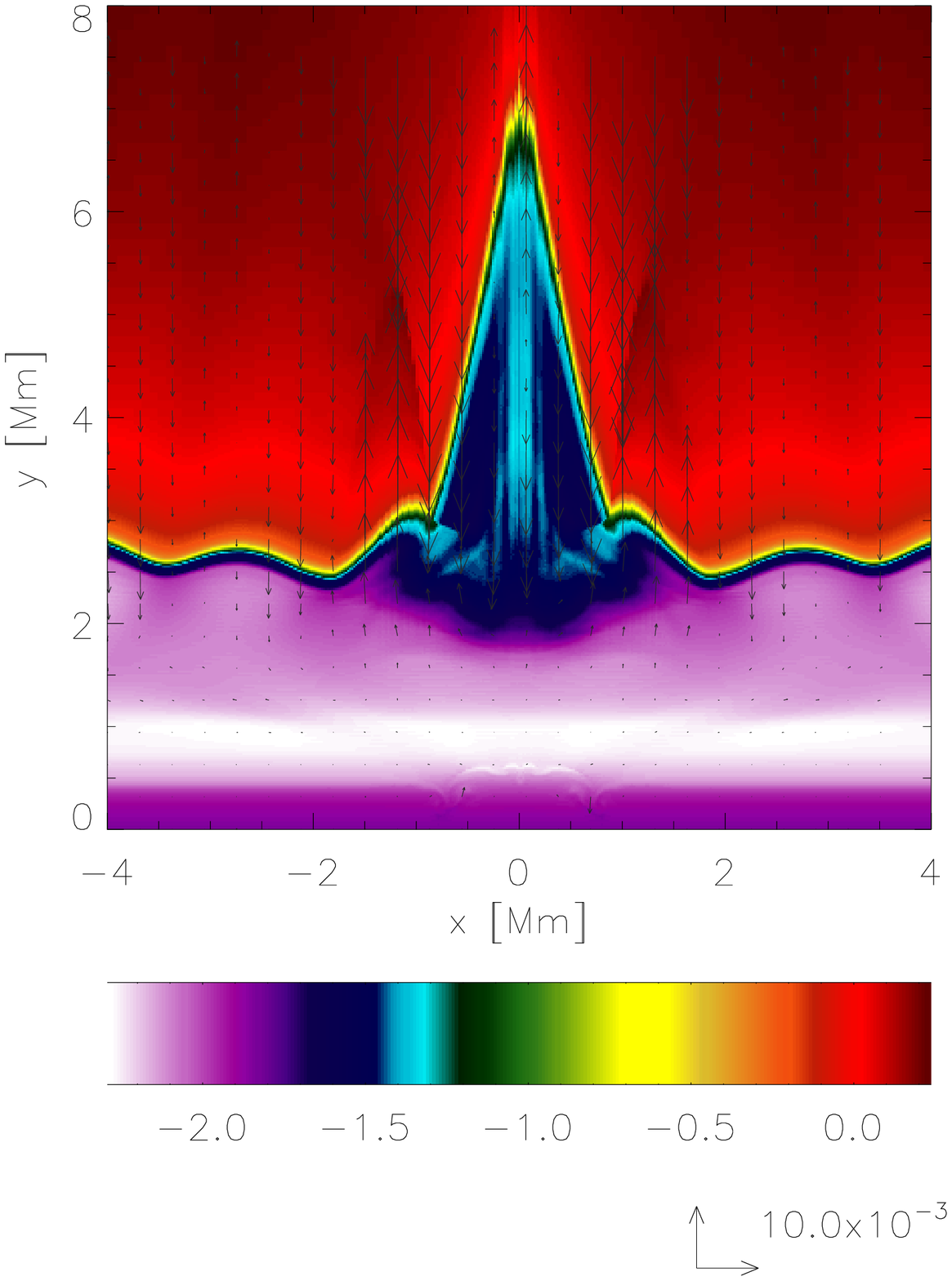}
\includegraphics[scale=0.31, angle=0, trim=0cm 0cm 0cm 0cm, clip=true]{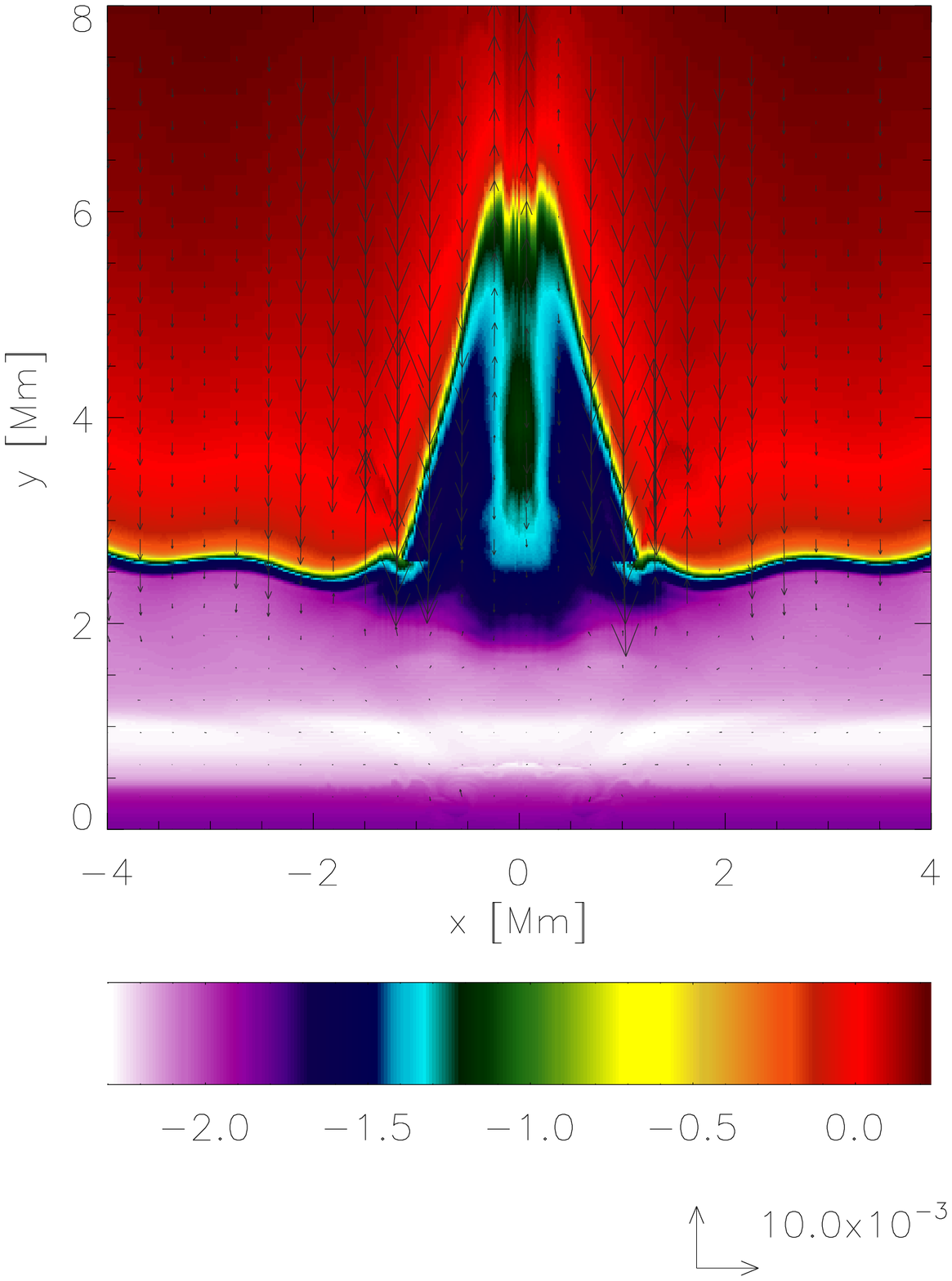}
\caption{\small 
Spatial profiles of logarithm of temperature (color maps) and velocity (arrows) profiles at 
$t=200$ s (top-left), $t=400$ s (top-right), 
$t=800$ s (bottom-left), $t=10^3$ s (bottom-right), 
for the case of the vertical equilibrium magnetic field. 
Temperature is expressed in units of $1$ MK. 
The arrow below each panel represents the length of the velocity vector, expressed in units of $10\times 10^{-3}$ Mm s$^{-1}$. 
{\bf The corresponding movie can be found in the file 
fig3.avi
online. }
}
\label{fig:By_T}
\end{figure}
First, we investigate the case of the vertical equilibrium magnetic field, which corresponds to ${\bf\hat s}={\bf\hat y}$ in Eq.~(\ref{eq:bb}). 
Here ${\bf\hat y}$ is a unit vector along the $y$-direction. The initial pulse of Eq.~(\ref{eq:perturb}) triggers magnetoacoustic-gravity waves in the solar atmosphere. 
These waves are compressive, and so we can trace their evolution in both temperature and velocity. Figure~\ref{fig:By_T} illustrates logarithm of temperature profiles 
(colour maps) and velocity vectors 
resulting from 
the perturbations. 
At $t=200$ s (top-left panel), the excited perturbations 
{\bf 
in the vertical component of velocity 
}
already penetrated the solar corona arriving to the point ($x=0$ Mm, $y=5.5$ Mm). 
{\bf 
The perturbation in the vertical component of velocity propagates by $\approx$$5$ Mm in first $200$ s of its evolution, 
which gives the speed of propagation
as $\approx$$25\times 10^{-3}$ Mm s$^{-1}$. The average sound speed in this region is significantly higher. 
This is also confirmed by Fedun et al. (2011b)
(cf., their Figs. 4-7) as their waves propagate from the transition region up by $2$ Mm within 
about $40$ s of the simulation time. From this comparison we infer that the perturbation we initially imposed 
triggers flows in the solar atmosphere first, and thereafter 
these flows produce magnetoacoustic-gravity waves. 
}
At this moment of time, the transition region experiences small perturbation and the signal reflected from the transition region is represented by 
arrows around $x=0$ Mm, $y=1.5$ Mm. This reflected signal clashes with the slowly up-going wave that results at the launching place, i.e., $x=0$ Mm, $y=0.5$ Mm. 
As a result of this clash, the leading front is generated. 
Cold chromospheric plasma lags behind 
and 
at $t=400$ s (Figure~\ref{fig:By_T}, top-right panel) it reaches an altitude of $y\simeq 6.5$ Mm. 
The reason for the cold material being lifted is the pressure gradient which works against gravity and 
forces the chromospheric material to penetrate the solar corona (Sterling \& Hollweg, 1989).
At $t=800$ s (Figure~\ref{fig:By_T}, bottom-left), the transition region experiences well-developed oscillations with the central region reaching the level of $y\simeq 7$ Mm at $x=0$ Mm,
and also the off-side propagating oscillations. These oscillations are of lower amplitudes at $t=10^3$ s (Figure~\ref{fig:By_T}, bottom-right panel). At this time, the central plasma experiences 
gravitational fall-off while the region of cold plasma widens. The jet, which is associated with the shock, lasts for several hundred seconds and its velocity reaches up to maximum
limit of $\sim 60\times 10^{-3}$ Mm s$^{-1}$. 

%
\begin{figure}[h]
\begin{center}
\includegraphics[scale=0.4, angle=0]{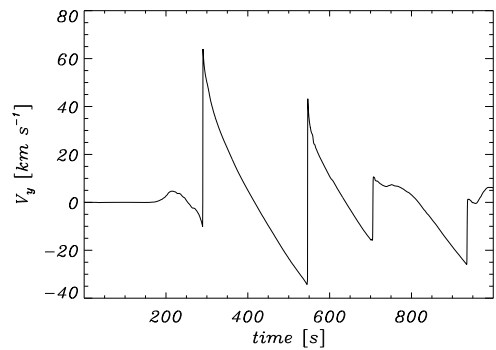}
\caption{\small
The time signature of $V_{\rm y}$ 
collected at the point $(x=0, y=5)$ Mm for the case of the vertical equilibrium magnetic field. 
}
\label{fig:By_ts}
\end{center}
\end{figure}
\begin{figure}[h]
\begin{center}
\includegraphics[scale=0.35, angle=90, trim=3cm 0cm 0cm 0cm, clip=true]{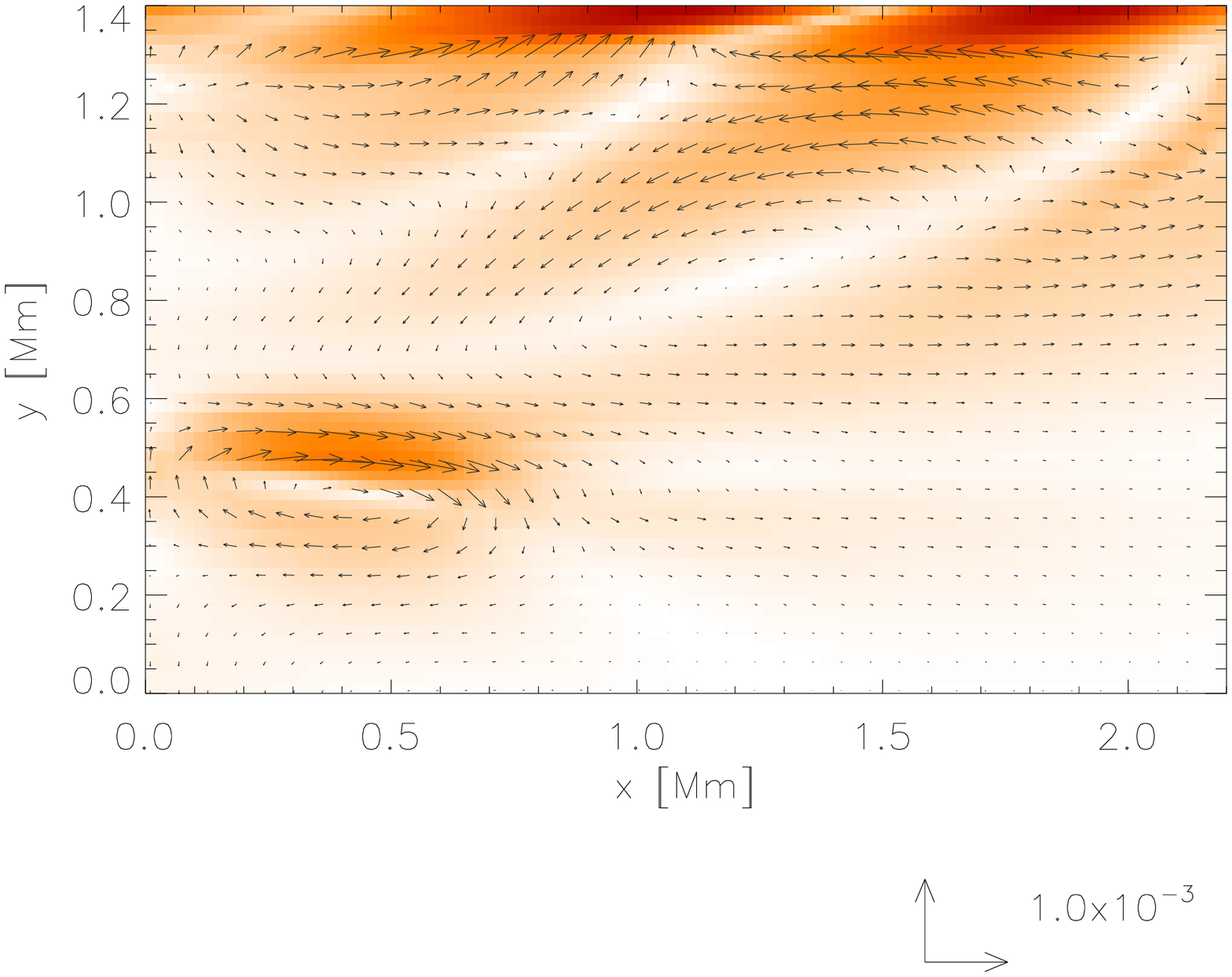}
\includegraphics[scale=0.35, angle=90, trim=0cm 0cm 0cm 0cm, clip=true]{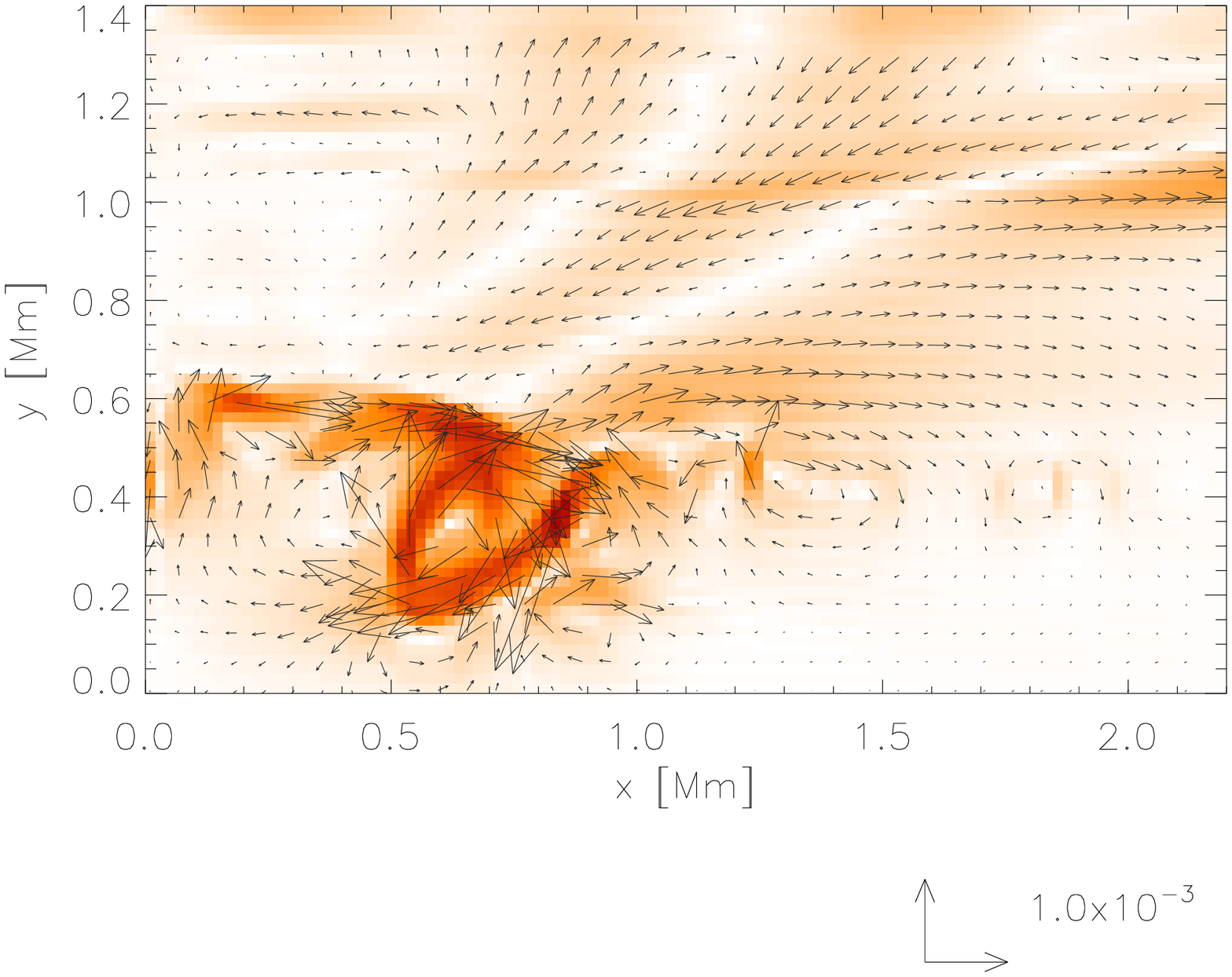}
\caption{\small
Total velocity, $\vert{\bf V}\vert$, (color map) and velocity (arrows) profiles at $t=400$ s (top) and $t=10^3$ s (bottom) 
for the case of the vertical equilibrium magnetic field. 
The velocity vectors are expressed in units of $1\times 10^{-3}$ Mm s$^{-1}$. 
{\bf The corresponding movie can be found in the file 
fig5.avi online. }
}
\label{fig:By_V}
\end{center}
\end{figure}

Figure~\ref{fig:By_ts} illustrates the vertical component of velocity that is collected in time at the detection point $(x=0, y=5)$ Mm for the case of Fig.~\ref{fig:By_T}. 
As a result of rapid mass density fall off with height, upwardly-propagating waves grow in their amplitude and steepen rapidly to form shocks. 
The arrival of the first shock-front at the detection point is clearly seen at $t\simeq 290$ s. The second shock-front reaches the detection point at $t\simeq 540$ s, 
i.e. approximately $250$ s later. 
This secondary shock (and subsequent shocks which arrive respectively after about $160$ s and $240$ s) results 
from the nonlinear wake, which lags behind the leading shock (Sterling \& Hollweg, 1989; Srivastava \& Murawski, 2011).
These times can be compared with the acoustic cut-off period (Roberts, 2006),
\beqa
\label{eq:P_ac}
P_{\rm ac}(y) = \frac{4\pi\Lambda(y)}{c_{\rm s}(y)\sqrt{1+2\Lambda'(y)}}\, ,
\eeqa
which for $y=0.5$ Mm attains a value of $P_{\rm ac}(y) \simeq 180$ s. 
This value 
differs by $70$ s from 
the time-span between arrivals of neighboring shocks, which is $250$ s. 
{\bf{However, the 2D model we discuss here is more complex than the 1D scenario which is described by the Klein-Gordon equation 
(Roberts, 2006)}}.
The wave-period we detected is altered by the interaction between up-going waves from the launching place of the initial pulse and 
the reflected one from the transition region signals. 
As wave reflections result at the large temperature gradients, therefore, the chromosphere sustains a cavity for these waves which are represented by the oblique stripes located at altitudes $0.7\, {\rm Mm} < y < 1.4\, {\rm Mm}$ (Fig.~\ref{fig:By_V}). 
In the neighborhood of the point $x=0.4\, {\rm Mm},\:y=0.4\, {\rm Mm}$, we observe the formation of vortices (Fig.~\ref{fig:By_V}, top panel) which 
experience energy cascade into smaller scales. These vortices are present till the end of our simulation runs (Fig.~\ref{fig:By_V}, bottom panel). 
The first vortex results from the initial pulse in $V_{\rm y}$, which is a characteristic feature of velocity perturbations. 
This first vortex seeds convection in the convectively unstable plasma layers. According to the Schwarzschild's instability condition, a medium is convectively unstable 
if the squared {\sl buoyancy} (or Brunt-V\"ais\"al\"a) 
{\sl frequency},
%
\beq
\omega_g^2\equiv g\left(\frac{1}{\gamma\Lambda}-\frac{1}{\Lambda}\right) = 
\frac{g^2}{c_s^2} - \frac{g}{\Lambda}\, ,
\eeq
%
is negative (e.g., Roberts, 2006). 
As this criterion is satisfied for $y < 0.6$ Mm, therefore, the convection sets in there. 
Such vortices were also theorized by 
Konkol, Murawski, \& Zaqarashvili (2011)
in a similar context. 

%
%
%
%
\subsection{Horizontal equilibrium magnetic field:  ${\bf\hat s}={\bf\hat x}$}\label{sub:horB}
\begin{figure}[h]
\begin{center}
\includegraphics[scale=0.315, angle=0]{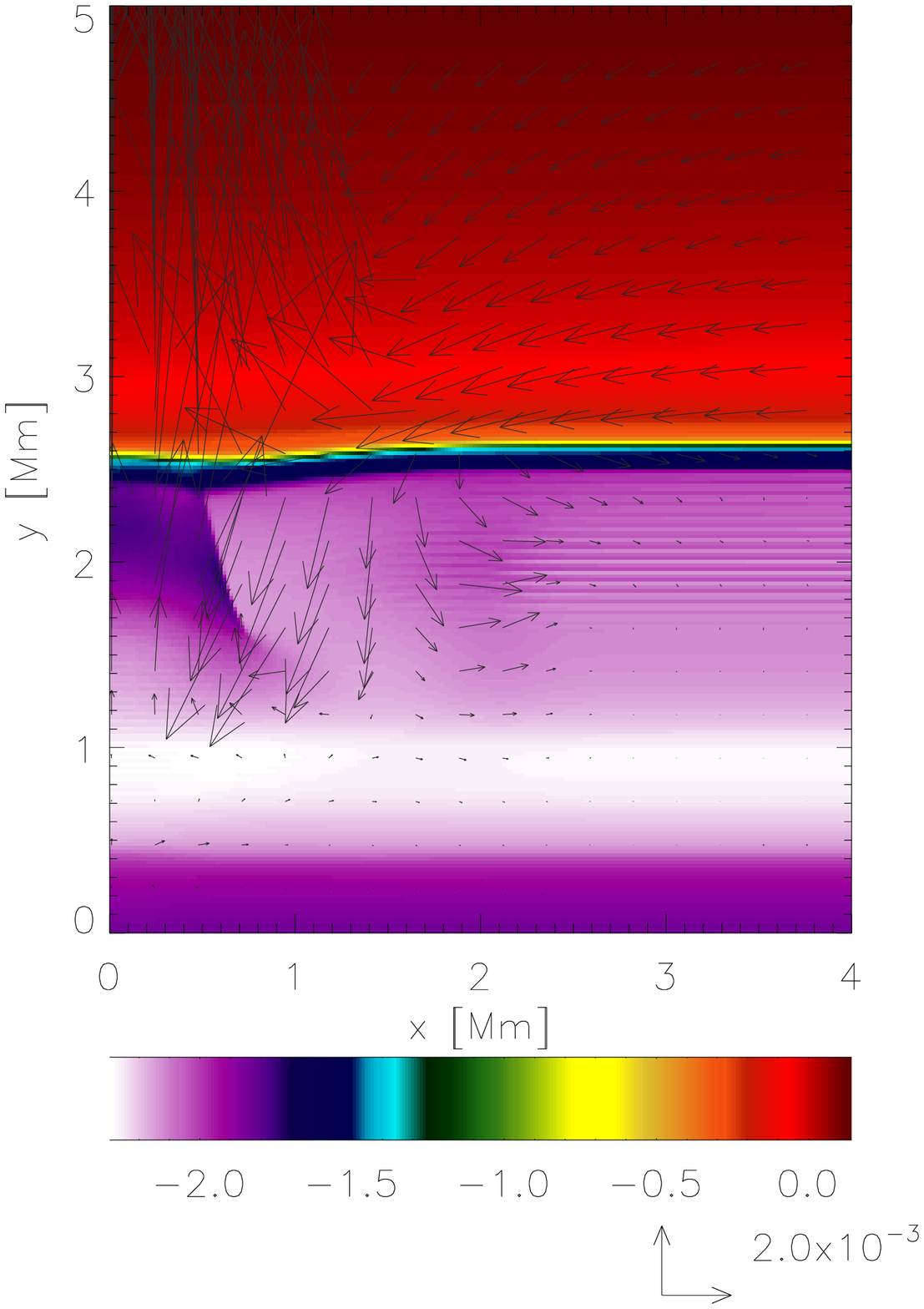}
\includegraphics[scale=0.315, angle=0]{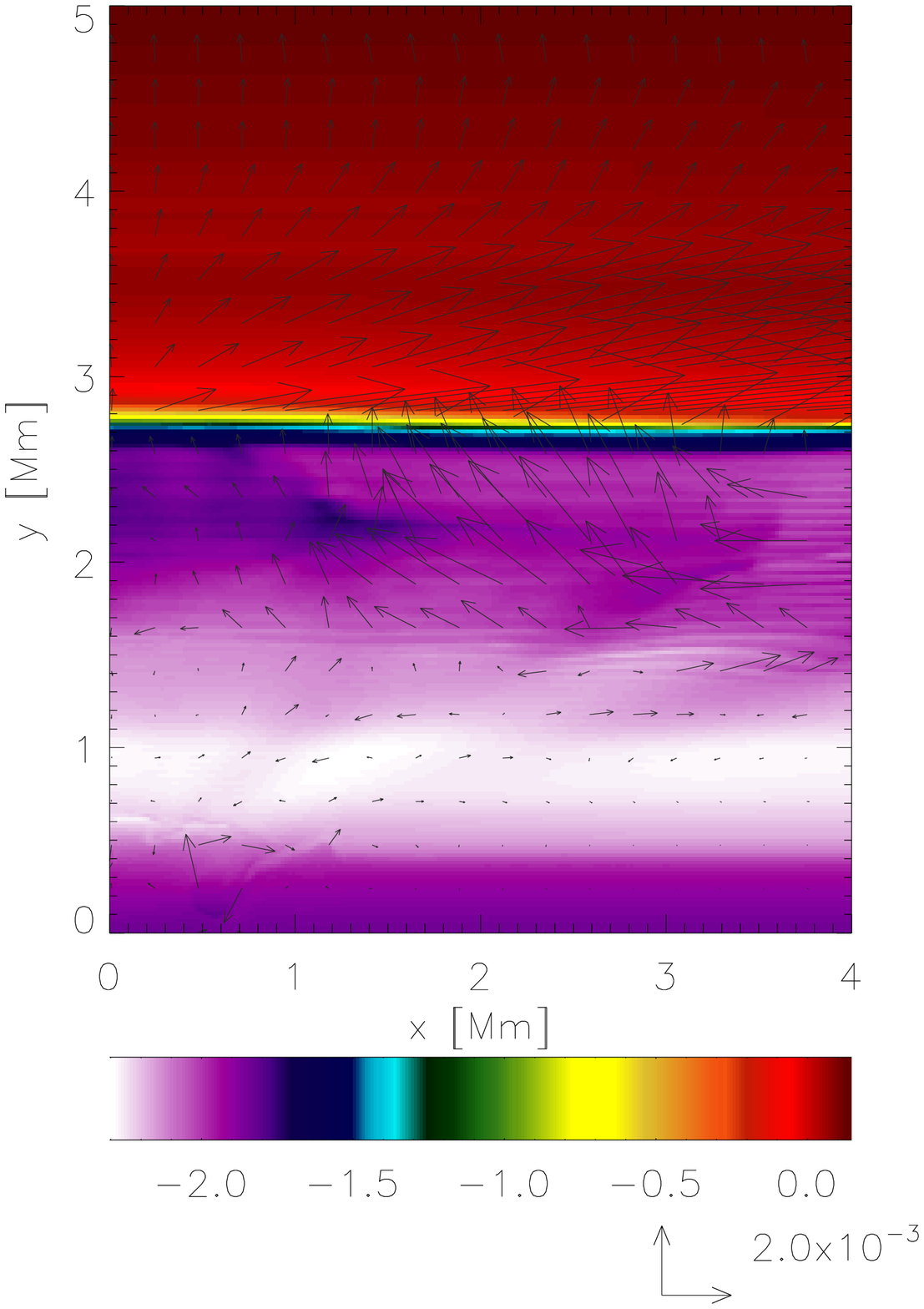}
\caption{\small 
Spatial profiles of logarithm of temperature (color maps) and velocity (arrows) at $t=250$ s (top) and $t=10^3$ s (bottom) 
for the case of the horizontal equilibrium magnetic field. 
Temperature is drawn in units of $1$ MK. The arrow below each panel represents the length of the velocity vector, expressed in units of 
$2\times 10^{-3}$ Mm s$^{-1}$. 
{\bf The corresponding movie can be found in the file 
fig6.avi online. }
}
\label{fig:Bx_T}
\end{center}
\end{figure}
In this section, we investigate the  horizontal equilibrium magnetic field, which corresponds to ${\bf\hat s}={\bf\hat x}$ in Eq.~(\ref{eq:bb}). 
In this case, the spatial profiles of $T$ drawn at $t=250$ s and $t=10^3$ s (Fig.~\ref{fig:Bx_T}) reveal small amplitude oscillations of the transition region 
without the presence of a jet which was observed for the case of the vertical magnetic field. On comparing with Fig.~\ref{fig:By_T}, 
the waves resulting from the initial pulses already arrived to the solar corona at $t=250$ s (Fig.~\ref{fig:Bx_T}, left panel). 
As the magnetic field 
{\bf 
lines are 
}
horizontal, therefore, these waves are essentially 
fast magnetoacoustic-gravity waves in the region of low plasma $\beta$ that occurs for $y>2.6$ Mm (Fig.~\ref{fig:initial_profile}, bottom panel). 
It is interesting that the orientation of magnetic field plays very crucial role in determining 
the amplitude of oscillations of the transition region. 

%
\begin{figure}[h]
\begin{center}
\includegraphics[scale=0.4, angle=0]{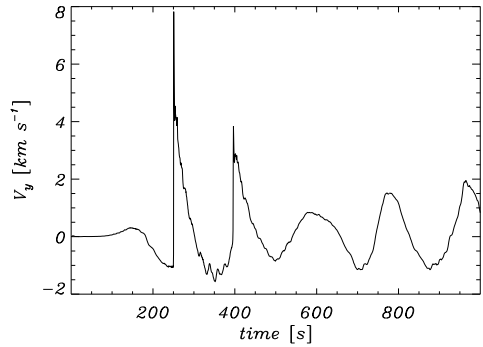}
\caption{\small
The time signature of $V_{\rm y}$ collected at the point $(x=0, y=5)$ Mm 
for the case of the horizontal equilibrium magnetic field. 
}
\label{fig:Bx_ts}
\end{center}
\end{figure}
Similar to the time-signatures of Fig.~\ref{fig:By_ts}, $V_{\rm y}$ collected at the detection point $(x=0,\: y=5)$ Mm, reveals the shocks. 
However, for the horizontal magnetic field, we now observe only two shocks (as seen in Fig.~\ref{fig:Bx_ts}). 
The first shock arrives to the detection point at $t=250$ s and the second shock reaches this point at $t=400$ s. 
We note that the time-span of $150$ s between these events is $100$ s shorter than in the case of the vertical magnetic field in 
Sect.~\ref{sub:vertB}. 
However, they are shorter only by $30$ s than the acoustic cut-off period, $P_{\rm ac}$ resulting from Eq.~(\ref{eq:P_ac}). 
Note that the subsequent three oscillations, which are of smaller amplitude, reveal waveperiods of the order of $P_{\rm ac}$. 

\begin{figure}[h]
\begin{center}
\includegraphics[scale=0.35, angle=90, trim=3cm 0cm 0cm 0cm, clip=true]{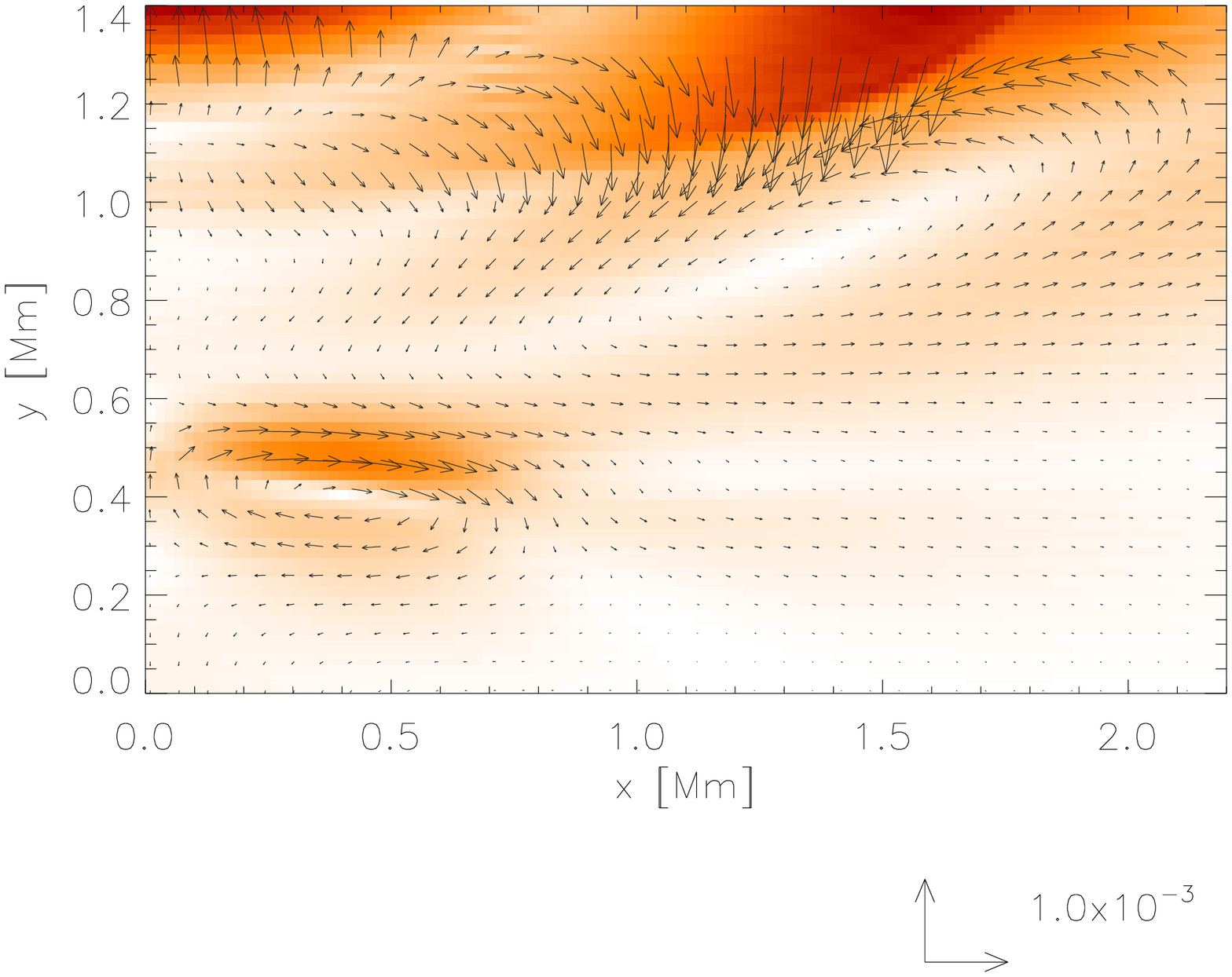}
\includegraphics[scale=0.35, angle=90, trim=0cm 0cm 0cm 0cm, clip=true]{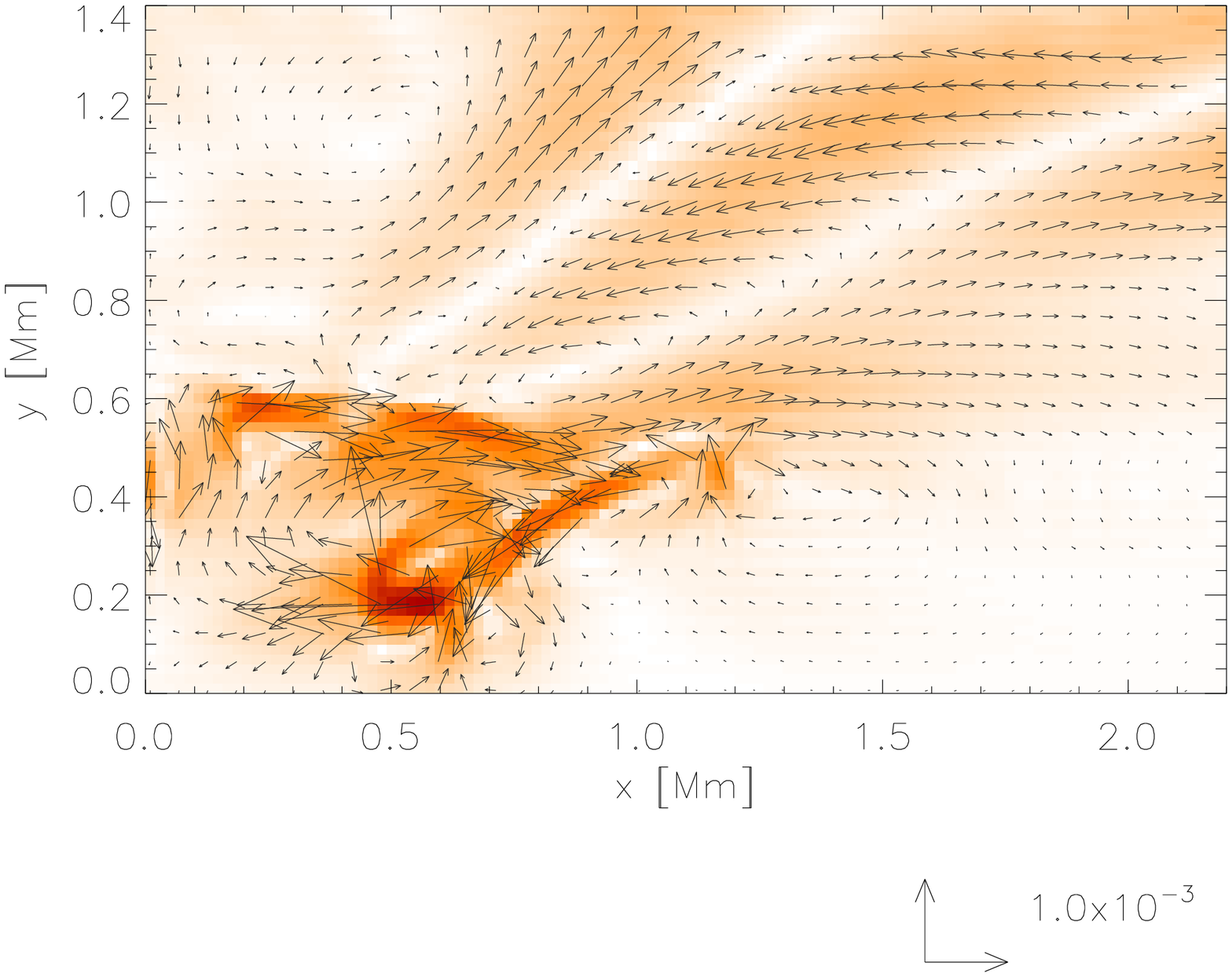}
\caption{\small
Total velocity, $\vert{\bf V}\vert$, (color map) and velocity (arrows) profiles at $t=400$ s (top) and $t=10^3$ s (bottom) 
for the case of the horizontal equilibrium magnetic field. 
The velocity vectors are expressed in units of $1\times 10^{-3}$ Mm s$^{-1}$. 
{\bf The corresponding movie can be found in the file 
fig8.avi online. }
}
\label{fig:Bx_V}
\end{center}
\end{figure}

The chromosphere exhibits essentially similar features to those seen for 
the vertical magnetic field of Sect.~\ref{sub:vertB}. Both, the vortex motion at $x=0.4\, {\rm{Mm}},\:y=0.4\, {\rm Mm}$ and 
trapped waves in the solar chromosphere for $0.7\, {\rm Mm}\, < y < 1.4\, {\rm Mm}$, are again discernible (Fig.~\ref{fig:Bx_V}). 
At $t=400$ s (top panel), the vortex is well developed. This vortex is similar to that of Fig.~\ref{fig:By_V} (top panel). 
The flow patterns at the line $y=1.4$ Mm 
in Figs.~\ref{fig:By_V} and \ref{fig:Bx_V} differ in some details though. 
In the case of the vertical magnetic field, the resultant waves are essentially slow magnetoacoustic-gravity waves, while Fig.~\ref{fig:Bx_V} 
corresponds to the horizontal background magnetic field with much contribution from fast magnetoacoustic-gravity waves. 
At a later time, turbulence results from this vortex and it differs in 
some details from its analog of Fig.~\ref{fig:By_V} (bottom panel). 
In both the vertical and horizontal magnetic fields, the turbulence originates from the impulsively triggered perturbations 
that are initially launched in the convectively unstable atmospheric layers. 

%
%
%
%
%
\subsection{Oblique equilibrium magnetic field:  ${\bf\hat s}= \left({\bf\hat x}+ {\bf\hat y}\right) / \sqrt{2} $}\label{sub:oblB}
\begin{figure}
\includegraphics[scale=0.31, angle=0]{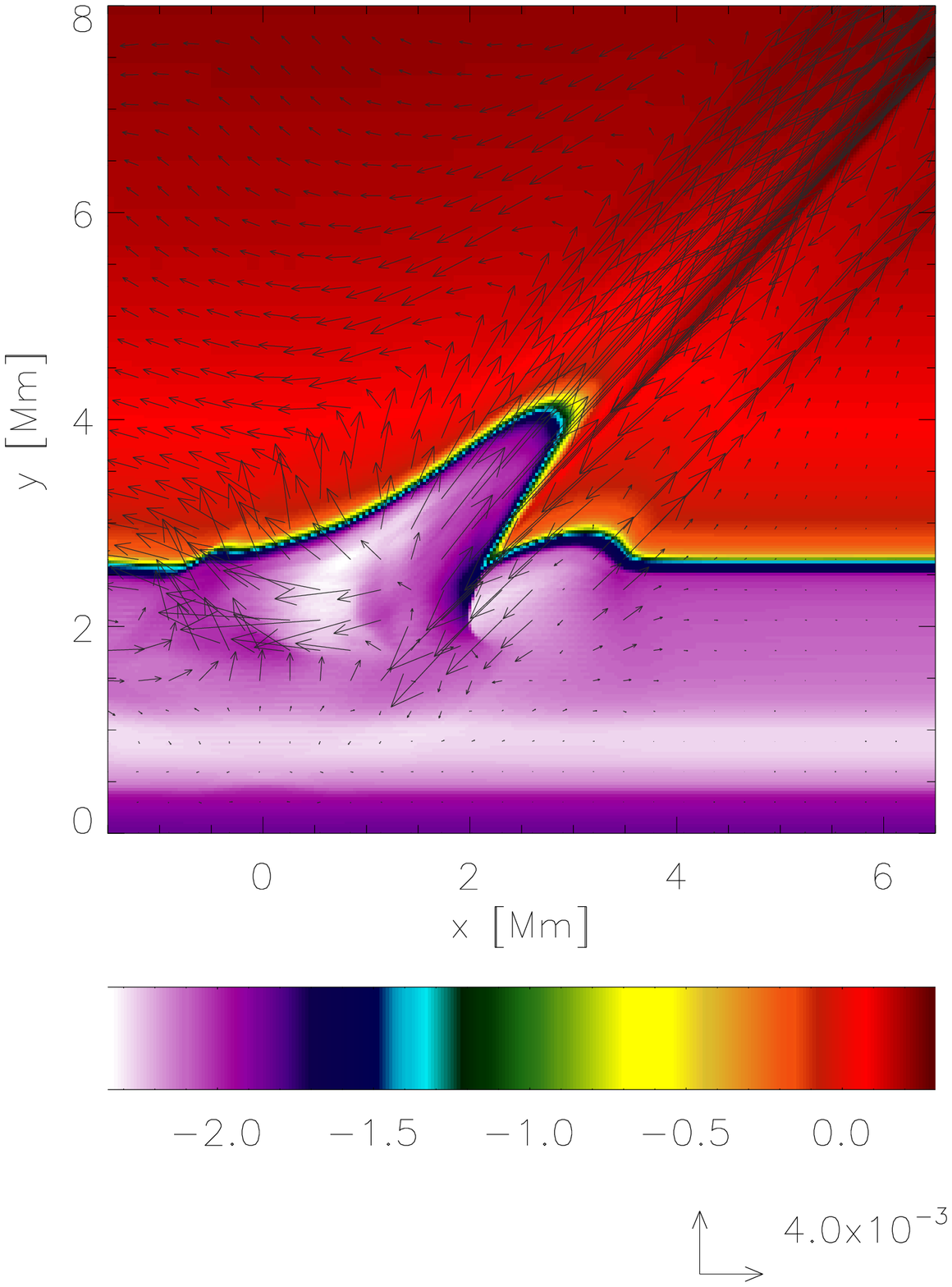}
\includegraphics[scale=0.31, angle=0]{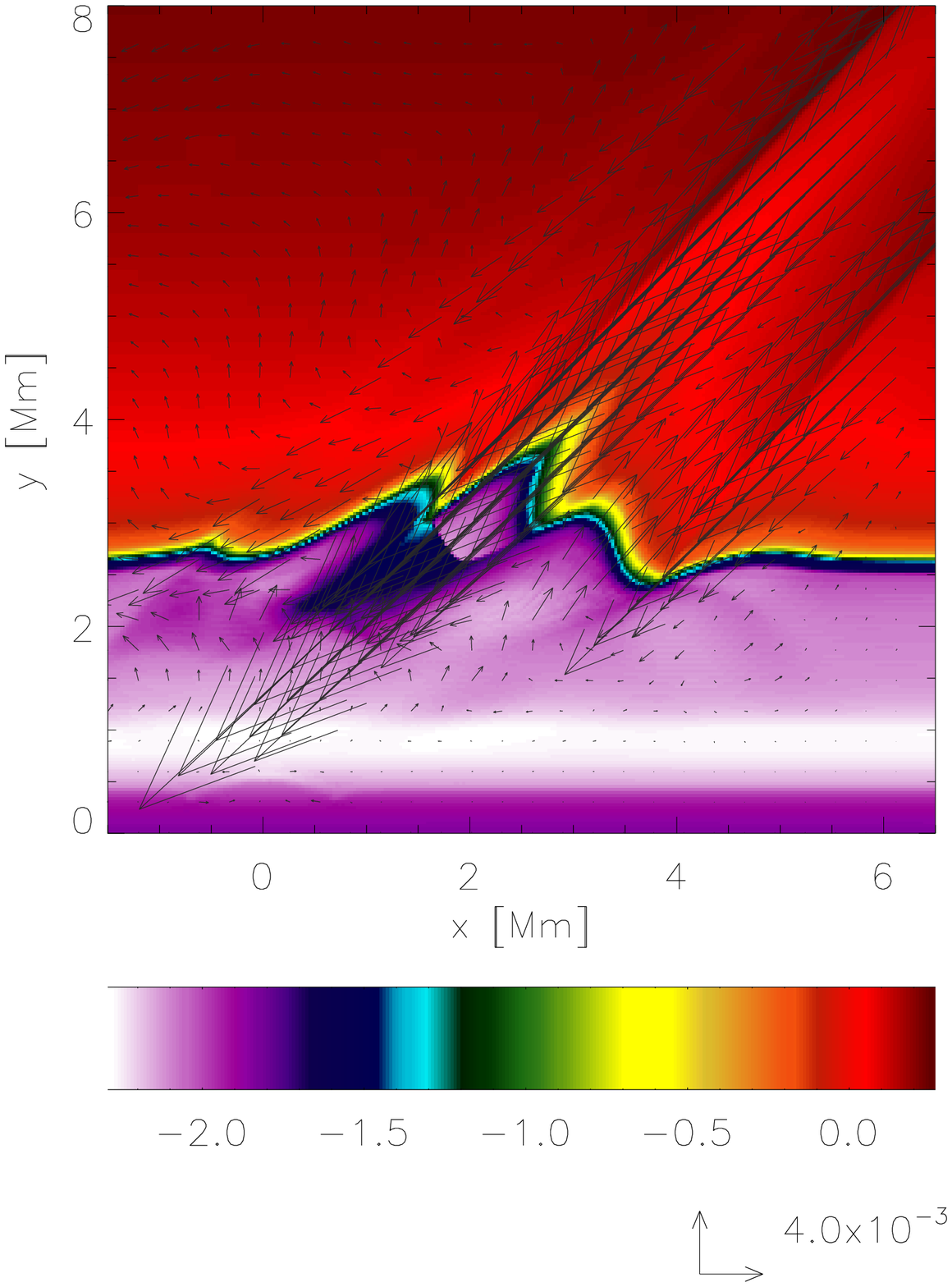}\\
\includegraphics[scale=0.31, angle=0]{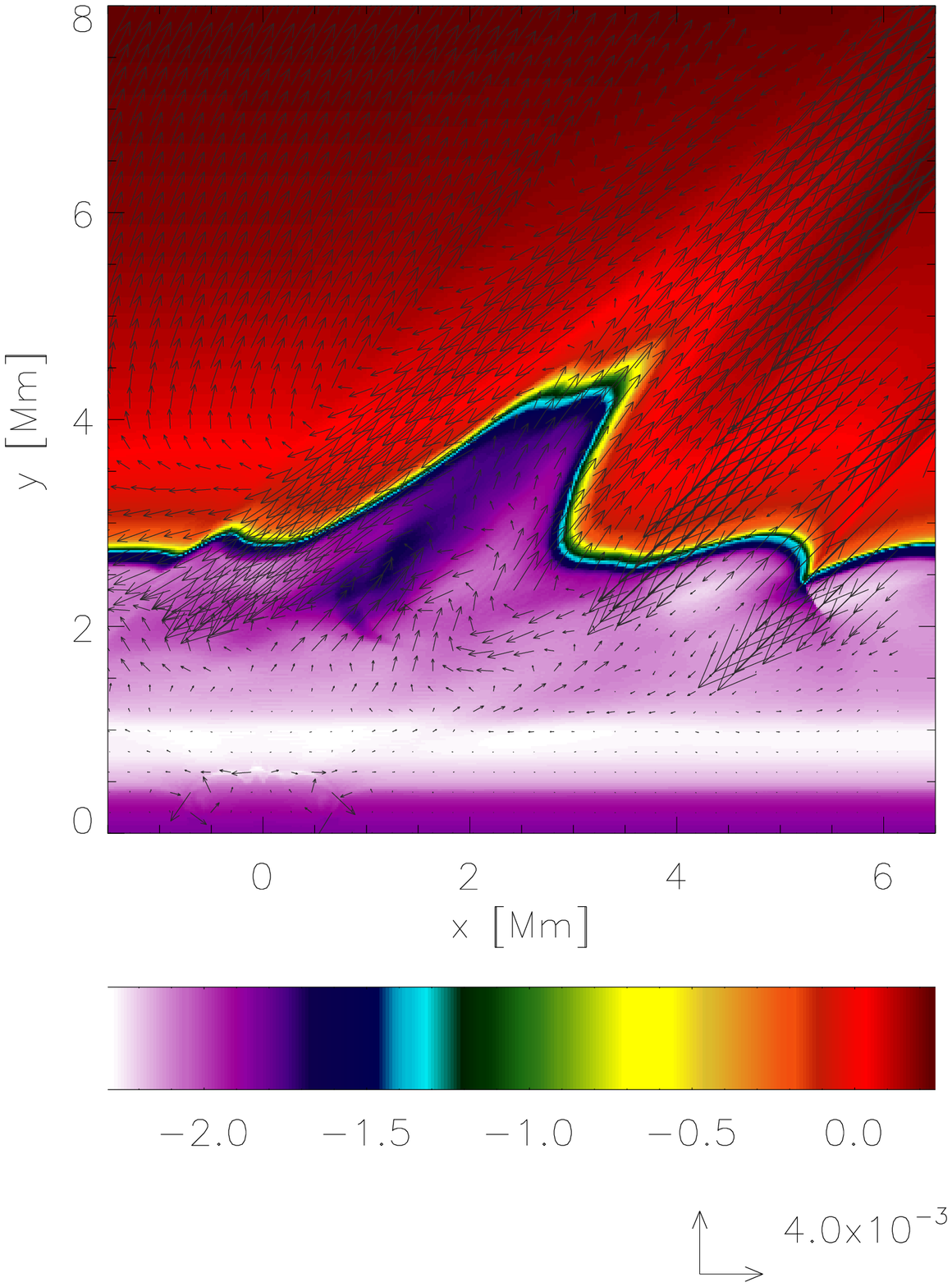}
\includegraphics[scale=0.31, angle=0]{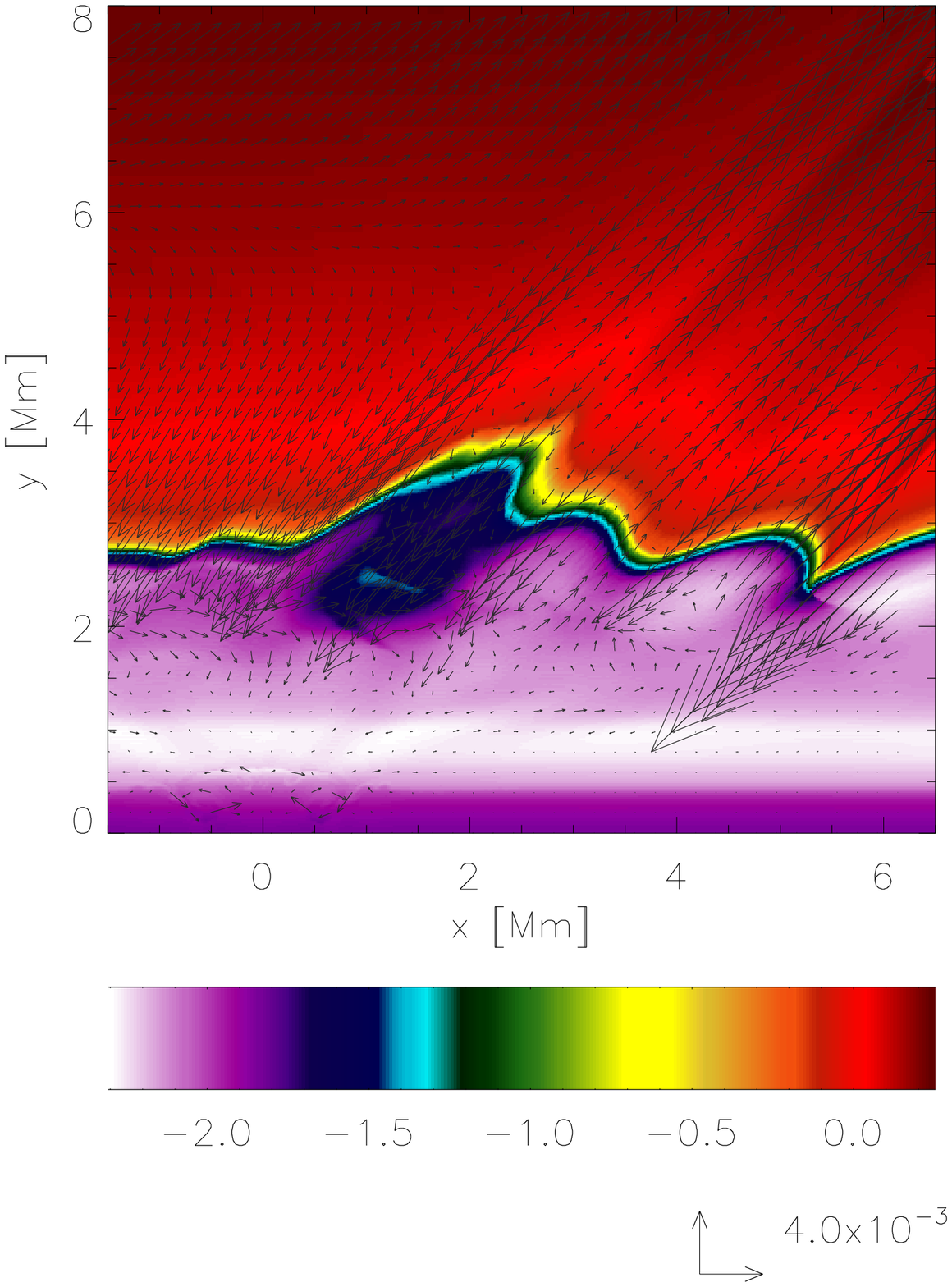}
\caption{\small 
Spatial profiles of logarithm of temperature (color maps) and velocity (arrows) profile at 
$t=400$ s (top-left), $t=600$ s (top-right), 
$t=800$ s (bottom-left), $t=10^3$ s (bottom-right)
for the case of the oblique equilibrium magnetic field. 
Temperature is drawn in units of $1$ MK. 
The arrow below each panel represents the length of the velocity vector, expressed in units of $4\times 10^{-3}$ Mm s$^{-1}$. 
{\bf The corresponding movie can be found in the file 
fig9.avi online. }
}
\label{fig:BxBy_T}
\end{figure}

\begin{figure}
\begin{center}
\includegraphics[scale=0.4, angle=0]{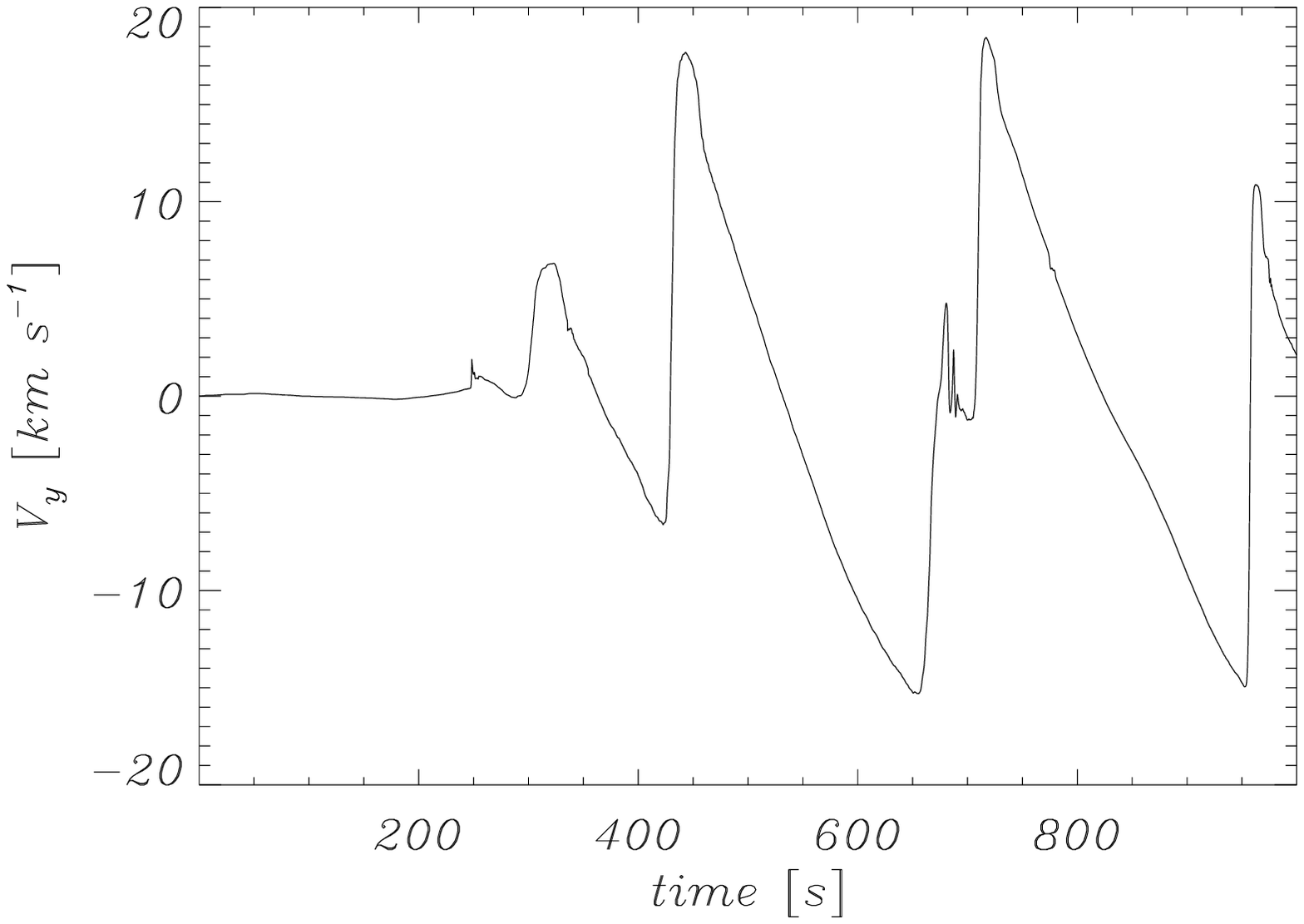}
\includegraphics[scale=0.4, angle=0]{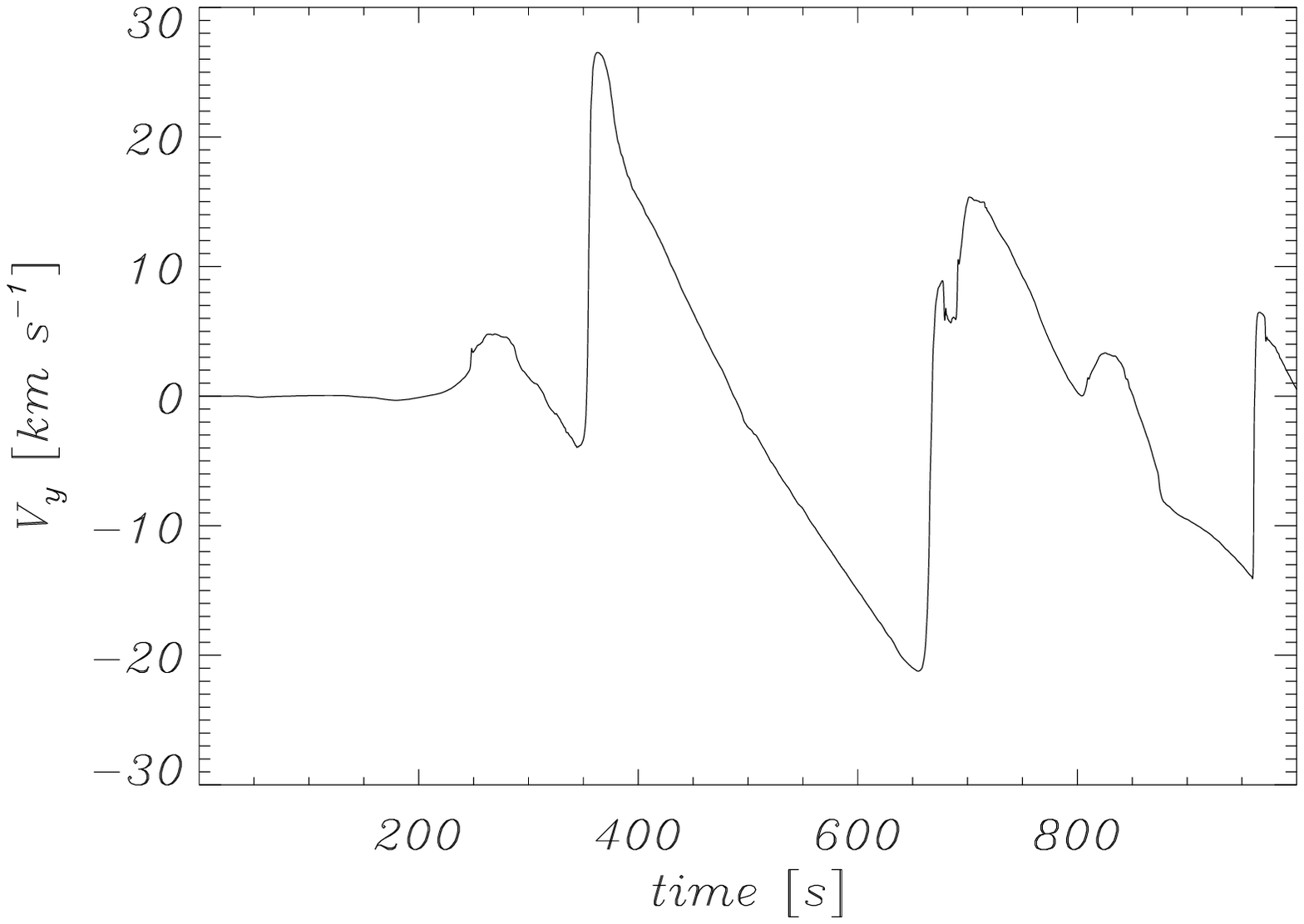}
\caption{\small
The time signature of $V_{\rm y}$ collected at the points 
$(x=4.5, y=5)$ Mm (top) and $(x=3.84, y=4.98)$ Mm (bottom) 
for the case of the oblique equilibrium magnetic field. 
}
\label{fig:BxBy_ts}
\end{center}
\end{figure}

\begin{figure}
\begin{center}
\includegraphics[scale=0.475, angle=90]{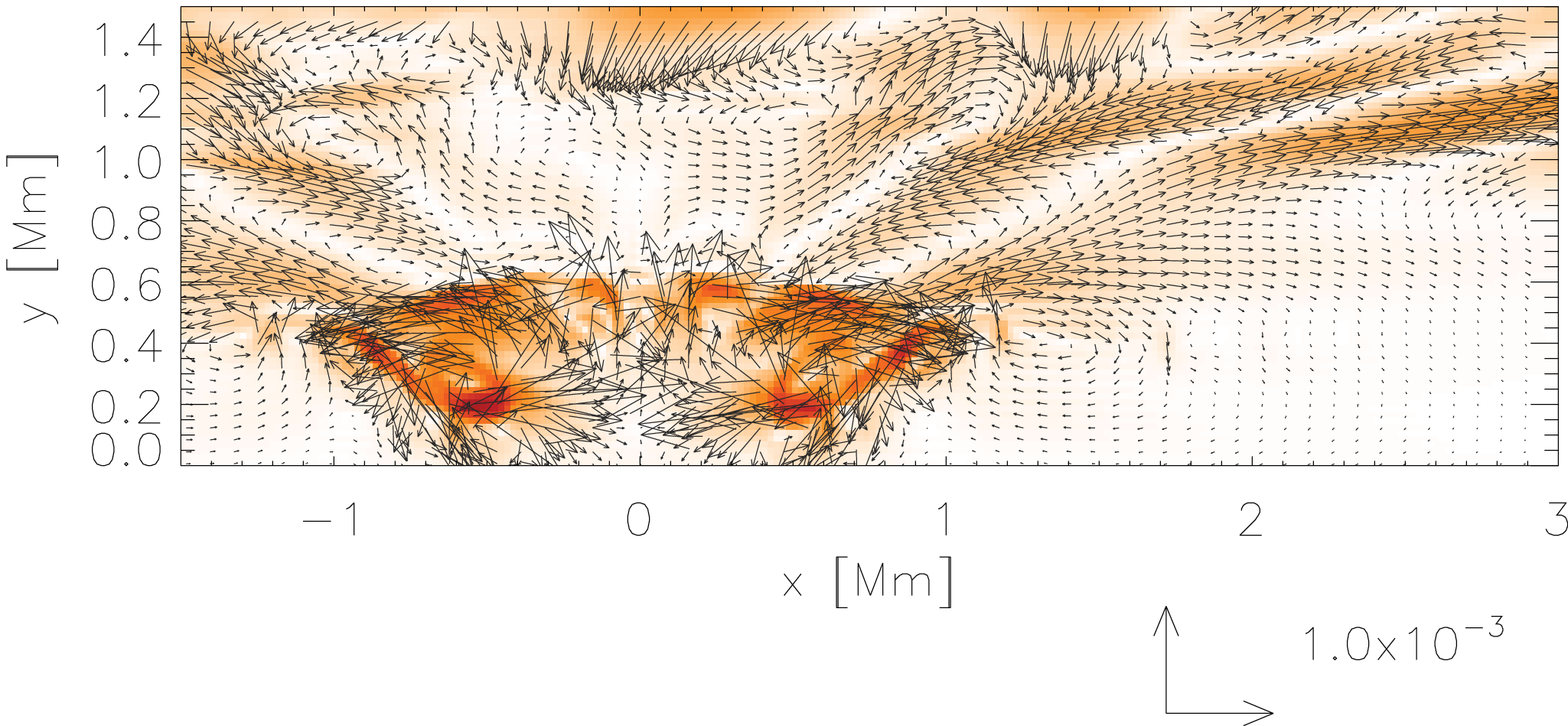}
\caption{\small
Total velocity, $\vert{\bf V}\vert$, (color map) and velocity (arrows) profile at $t=10^3$ s
for the case of the oblique equilibrium magnetic field. 
The velocity vectors are expressed in units of $1\times 10^{-3}$ Mm s$^{-1}$. 
{\bf The corresponding movie can be found in the file 
fig11.avi online. }
}
\label{fig:BxBy_V}
\end{center}
\end{figure}

In this section, we discuss 
the case of the oblique magnetic field, for which ${\bf\hat s}$ in Eq.~(\ref{eq:bb}) is at an angle of $\pi/4$ to the horizontal axis. 
Figure~\ref{fig:BxBy_T} displays temperature profiles at $t=400$ s (top-left panel), $t=600$ s (top-right panel), $t=800$ s (bottom-left panel) and $t=10^3$ s (bottom-right panel). 
These profiles are more complex than those for either the horizontal or the vertical magnetic field of Sect.~\ref{sub:vertB} and 
Sect.~\ref{sub:horB}. 
At $t=400$ s (Figure~\ref{fig:BxBy_T}, top-left), we clearly see the rising transition region. At $t=10^3$ s (Figure~\ref{fig:BxBy_T}, bottom-right), 
we observe surface waves propagating along the transition region. However, for the oblique equilibrium magnetic field, we find that time signatures of $V_{\rm y}$ 
reveal shocks, 
and the waveperiods are about $250-300$ s (Fig.~\ref{fig:BxBy_ts}). 
The jet, which is associated with the shock, lasts for several hundred seconds and its velocity reaches up to maximum
limit of $\sim 30\times 10^{-3}$ Mm s$^{-1}$. 
This is a result of complex interaction of waves with the background plasma in upper regions of the solar atmosphere. This complexity is seen at $t=10^3$ s in the velocity profiles of Fig.~\ref{fig:BxBy_V}, 
which clearly illustrates the trapped and reflected waves in the solar chromosphere as well as well-developed vortices. 
%


%
%


%
%


%
\section{Discussion and conclusion}\label{SECT:DISS}
%
We have investigated the impulsive excitation of magnetoacoustic-gravity oscillations 
and have compared and contrasted their resultant propagation under different orientations of equilibrium magnetic fields. 
We performed 2D numerical simulation of the velocity and gas pressure pulses, which mimic a solar granule. These pulses were initially launched at the top of solar photosphere, 
in a stratified solar atmosphere utilizing the VAL-C temperature profile. 
We find that in the cases the background magnetic field possesses a non-zero vertical component 
the amplitude of the upwardly-propagating perturbations rapidly grows with height due to the rapid decrease in the equilibrium mass density. 
Therefore, the perturbation quickly steepens into shocks in the upper regions of the solar chromosphere that launches the cool material behind it. 


We find that the excitation of magnetoacoustic-gravity waves in a non-ho\-ri\-zon\-tal equilibrium magnetic field configuration 
is channeled upwards and is able to cause large amplitude transition region oscillations. 
Meanwhile, the horizontal equilibrium magnetic field configuration results in comparatively smaller amplitude transition region oscillations, 
due to fast magnetoacoustic-gravity waves which spread their energy across magnetic field lines in contrast to slow magnetoacoustic-gravity waves which in a strongly magnetized plasma region 
are guided along magnetic field lines. 
This indicates that the wave energy is transfered into the upper atmosphere in larger amounts where the magnetic field is more vertical. 
However, the complexity and therefore 
the evolution of 
a horizontal equilibrium magnetic field 
allows the reflection and trapping of the waves in the lower solar atmosphere and can influence 
the localized dynamics and heating of the atmosphere. The reflection of the waves in the lower solar atmosphere and its trapping 
may also result in the presence of chromospheric cavities. 

By analyzing the time-signature of $V_{\rm{y}}$ collected at a fixed spatial point, we also find that the period of oscillation is {\emph{longer}} 
when comparing the vertical magnetic field case (here, the characteristic period of oscillation was $\approx 250-300$ s) with 
the horizontal equilibrium magnetic system (in which the characteristic period of oscillation was $\approx 150-200$ s). 
For the vertical magnetic field system in low plasma $\beta$ regions, the magnetoacoustic-gravity waves are 
well-described as {\emph{slow}} magnetoacoustic-gravity waves (propagating along the magnetic field lines at approximately the sound speed, $c_{\rm s}(y)$, see Eq.~\ref{eq:c_s}). 
For the horizontal magnetic field system in the strongly magnetized plasma, the oscillations transverse to the magnetic field are well-described as 
{\emph{fast}} magnetoacoustic-gravity waves (propagating across the magnetic field lines at 
the fast speed, $c^2_{\rm f}(y) = c_{\rm A}^2(y) + c_{\rm s}^2(y)$, see Eqs.~\ref{eq:c_A} and \ref{eq:c_s}). 
As for $y=0.5$ Mm we have $c_{\rm A}\ll c_{\rm s}$, therefore the cut-off frequencies of the fast and slow magnetoacoustic-gravity waves are very close to each other, which results in 
very close values of the cut-off waveperiods. Therefore, we could expect that the detected waveperiods 
for the vertical and horizontal magnetic fields 
exhibit similar values. 
However, the cut-off waveperiods are derived on the basis of the linear theory which is valid for small amplitude oscillations only. 
As such small oscillations are present in the case of the horizontal magnetic field, therefore, the obtained numerical data lies close to the analytical prediction. 
In the case of the vertical magnetic field, the upwardly propagating waves interact with the waves which become reflected from the transition region. 
As a result of that larger amplitude oscillations originate, which significantly alter the background plasma. The waves, which propagate through 
this strongly modified medium exhibit modified velocities and they get reflected from the transition region that is locally largely curved. 
As a consequence of that waveperiods within the range $250-300$ s result, which differ from $P_{\rm ac}$. 


{\bf It is known that radiation is an effective mechanism of wave damping in the low photosphere ((Mihalas \& Toomre, 1982).
It might not radically alter the system's behavior, but radiative damping is at least likely to reduce the amplitude of the waves reaching the transition region, 
leading to shorter jets than what we see in our models, as well as lower amplitudes of the coronal shocks. 
In this paper, we do not invoke the radiative cooling nor thermal conduction in our model atmosphere as we aim to model 
the small-scale atmospheric regions above the solar photosphere where such effects are not believed to be dominant. However, we intend to include these effects in our future studies. 

It is noteworthy that the solar atmosphere is structured by convective overshoot which is absent in the model we devised. 
Instead, we isolated a single granule-like perturbation by aiming to mimic a magnitude of flow and plasma temperature 
that are associated with the solar granulation (Baran 2011), 
and considered the complex scenario which results by this simple model
to understand the physics of wave phenomena above such magnetic structures in the solar atmosphere. 
We intend to develop more advanced models in our future studies.}

In conclusion, our numerical simulations clearly demonstrate that 
small amplitude initial pulses in vertical velocity and gas pressure are 
able to trigger a plethora of dynamic phenomena in the upper regions of the solar atmosphere 
with waveperiods within the range of $150-300$ s, a value which depends on orientation of the background magnetic field. 
However, it should be noted that the performed 2D simulations are idealized in the sense that 
they do not include radiative transfer and thermal conduction along field lines. 
The magnetic field configuration and the equilibrium stratification are simple and we modeled a single granule only. 
These limitations require additional studies which we intend to carry on in near future. 
%
%
%

\begin{acks}
{\bf 
The authors express their thanks to the referee for their stimulating comments. 
}
This work has been supported by a Marie Curie International Research Staff Exchange Scheme Fellowship within the 7th
European Community Framework Program (K.M.). 
This research was carried out with the support of 
the "HPC Infrastructure for Grand Challenges of Science and Engineering" Project, co-financed by the European Regional Development Fund under the Innovative Economy Operational Program (K.M.). 
The software used in this work was in part developed by the DOE-supported ASC/Alliance Center for 
Astrophysical Thermonuclear Flashes at the University of Chicago. JM acknowledges IDL support provided by STFC, UK. 
AKS thanks Shobhna Srivastava for encouragements and supports. RO acknowledges financial support from MICINN/MINECO and FEDER funds
through grant AYA2011-22846 and also CAIB through the ``Grups Competitius'' scheme and FEDER funds. 
KM expresses his thanks to Kamil Murawski for his assistance in drawing the numerical data. 
\end{acks}

%
%


%


\end{document}